\documentclass[showpacs,aps,prd,nofootinbib,floatfix,amsmath,amssymb]{revtex4}
\usepackage{graphicx}
\usepackage{slashed}
\usepackage{mathtools}
\usepackage{multirow,array}
\usepackage{tikz}
\usetikzlibrary{decorations.markings}
\numberwithin{equation}{section}

\begin{document}

\makeatletter
\newbox\slashbox \setbox\slashbox=\hbox{$/$}
\newbox\Slashbox \setbox\Slashbox=\hbox{\large$/$}
\def\pFMslash#1{\setbox\@tempboxa=\hbox{$#1$}
  \@tempdima=0.5\wd\slashbox \advance\@tempdima 0.5\wd\@tempboxa
  \copy\slashbox \kern-\@tempdima \box\@tempboxa}
\def\pFMSlash#1{\setbox\@tempboxa=\hbox{$#1$}
  \@tempdima=0.5\wd\Slashbox \advance\@tempdima 0.5\wd\@tempboxa
  \copy\Slashbox \kern-\@tempdima \box\@tempboxa}
\def\FMslash{\protect\pFMslash}
\def\FMSlash{\protect\pFMSlash}
\def\miss#1{\ifmmode{/\mkern-11mu #1}\else{${/\mkern-11mu #1}$}\fi}
\makeatother

\title{Vacuum polarization in Yang-Mills theories with Lorentz violation}
\author{ J. J. Toscano and O. V\' azquez-Hern\' andez}
\affiliation{Facultad de Ciencias F\'isico Matem\'aticas, Benem\'erita Universidad Aut\'onoma de Puebla, Apartado Postal 1152 Puebla, Puebla, M\'exico.}
\begin{abstract}
The renormalizable extension of a pure Yang-Mills theory with Lorentz violation is characterized by the CPT-Even $(k_F)_{\mu \nu \lambda \rho}$ and the CPT-Odd $(k_{AF})_\mu$ constant Lorentz coefficients. In this paper, the one-loop structure of the theory up to second order in these Lorentz violating coefficients is studied using the BFM-gauge. Results for the diverse beta functions are derived and contrasted with those given in the literature at first order in these parameters. Special emphasis is putted on the beta function $\beta(g)$, which is studied in both mass-independent and mass-dependent renormalization schemes. It is found that in a mass-independent scheme the $(k_{AF})_\mu$ Lorentz coefficient does not contribute to the $\beta(g)$ function, but it does in a mass-dependent scheme with contributions that are gauge-dependent and IR divergent.
\end{abstract}


\maketitle
\section{Introduction}
\label{I} Special relativity is the essential building block in formulating theories of fundamental physics, especially in the small-distance regime where it is merged with quantum principles. However, there are well-founded reasons to suspects that Lorentz invariance is not an exact symmetry at the Planck length and below. Clues of Lorentz violation (LV) arise at a fundamental level from efforts to merge quantum theory and general relativity into a unified theory~\cite{QTvsGR}. Explicit or spontaneous LV has been studied in Planck-scale formulations, such as string theory~\cite{STLV}, noncommutative geometry~\cite{NCGLV}, loop quantum gravity~\cite{LQGLV}, and other contexts~\cite{OCLV}. Therefore, it is important to look for signs of LV at low energies. At low energies, the effects of Lorentz and CPT violation can be described in a model-independent way by the standard model extension (SME), which is an effective field theory that contains general relativity and the standard model (SM)~\cite{SMEGR,SME}. In its minimal version (mSME)~\cite{SME}, the model contains only renormalizable interactions, in the sense of mass units, but nonrenormalizable interactions are expected to play a dominant role at higher energies~\cite{NRSME,NR1,NR2,NR3,NR4}. The SME is an effective field theory, which does not involve new degrees of freedom, but is built from pieces that involve only the SM fields. Besides the SM part, the mSME Lagrangian includes a sum of all independent operators of up to dimension four of the form $T^{\mu_1 \mu_2 \cdots}{\cal O}_{\mu_1 \mu_2 \cdots}$, with $T^{\mu_1 \mu_2 \cdots}$ a constant Lorentz tensor and ${\cal O}_{\mu_1 \mu_2 \cdots}$ a Lorentz tensor which is gauge invariant and depends on SM fields. The $T^{\mu_1 \mu_2 \cdots}$ coefficients introduce LV, since they specify preferred directions in the spacetime. The criterion to construct an effective Lagrangian that incorporates LV is that it must be invariant under observer Lorentz transformations (coordinate transformations or passive transformations) but not under particle transformations (transformations of the experimental setup or active transformations)~\cite{OLT}. This must be so because two observers must agree with the result of a measurement regardless of whether Lorentz invariance is violated or not. Consequently, under this type of transformation both the ${\cal O}_{\mu_1 \mu_2 \cdots}$ operators and their coefficients $T^{\mu_1 \mu_2 \cdots}$ are recognized as Lorentz covariant objects of the same range. On the other hand, particle Lorentz transformations means transforming the experimental setup in such a way that an interaction of the background fields $T^{\mu_1 \mu_2 \cdots}$ with the devices can be detected through the measurements made before and after the transformation. This type of transformation only acts on the degrees of freedom of the theory, that is, the ${\cal O}_{\mu_1 \mu_2 \cdots}$ operators transform covariantly, but the $T^{\mu_1 \mu_2 \cdots}$ coefficients do not transform, so the mSME Lagrangian is not Lorentz invariant.\\

Experiments with stable particles, such as photons, electrons, and protons, have been used to investigate possible signs of LV. Experimental devises to investigate \textit{rotation} invariance and \textit{boost} invariance using photons have been designed for a long time. The most representative examples are the famous Michelson-Morley experiment~\cite{MME} and the Kennedy-Thorndike experiment~\cite{KTE}. While the former shows that the speed of the light is independent of the orientation of the apparatus, the latter shows that it is also independent of the velocity of the apparatus in different inertial frames. Some modern experimental devices have made it possible to establish severe limits on LV. Among others, we have a modern Michelson-Morley experiment that uses ultrastable oscillator frequency sources~\cite{MMME}, experiments with microwave resonators operating in Whispering Gallery modes~\cite{WG}, penning traps, which are devices  that use electric and magnetic fields to keep stable charged particles trapped for a long time~\cite{PT}, experiments with polarized electrons~\cite{PE}, and some noble gas maser~\cite{NGM}. Muon spectroscopy experiments have also been carried out~\cite{MUE}. Beyond terrestrial experiments, stringent bounds have derived from astrophysical sources~\cite{ASLV}.\\

In the literature, attention has focused on the first-order effects of $T^{\mu_1 \mu_2 \cdots}$ coefficients, since they describe possible experimental implications on Lorentz violation and also because these coefficients are expected to be quite suppressed. Although it should be mentioned that one-loop higher order effects have been studied in the context of the minimal quantum electrodynamics extension \cite{SecondQED,HigherQED}. On the other hand, to consider effects on SM observables, that is, observables that are invariant under both observer and particle Lorentz transformations, it is necessary to include second-order effects in the $T^{\mu_1 \mu_2 \cdots}$ coefficients. It is expected that first-order effects of $T^{\mu_1 \mu_2 \cdots}$ impact experiments designed to detect Lorentz violation, such as those aforementioned, but second-order effects of these coefficients can, in addition, contribute to SM observables, such as the static electromagnetic properties of elementary particles~\cite{OP1,OP2,OP3}. For instance, in the mSME the electromagnetic $\bar{f}f\gamma$
vertex, with $f$ stands for a lepton or quark, can develop at the one-loop level new electromagnetic structures proportional to the $T^{\mu_1 \mu_2 \cdots}$ coefficients~\cite{OP1} and new contributions that modify the usual ones if second order effects of the form $T^2=T^{\mu_1 \mu_2 \cdots}T_{\mu_1 \mu_2 \cdots}$ are considered~\cite{OP2,OP3}. Although these quantities do not carry information on spatial directions or relative motion, they can provide useful information about the importance of LV effects when constrained from high precision experiments, such as, for example, magnetic and electric dipole moments of charged leptons and nucleons. In particular, bounding these effects is particulary useful in cases of antisymmetric 2-tensors $T_{\mu \nu}=-T_{\nu \mu}$, since this types of objects are made of two spatial $\textbf{e}$ and $\textbf{b}$ vectors and thus bounds for $|\textbf{e}|$ and $|\textbf{b}|$ can be derived~\cite{OP2}. As we will see later, the study of second-order effects is valuable in itself since it can shed light on the very structure of the theory. This is the spirit of the present work.\\

Quantum field theories (QFTs) that violate Lorentz symmetry, such as the mSME, can lead to results that strongly contrast with those predicted by QFTs that preserve this symmetry. To illustrate this point, let us to comment the case of the static electromagnetic properties of spin-$\frac{1}{2}$ charged particles. It is a well-known fact that in the SM (and in any Lorentz-invariant QFT) the anomalous magnetic moment of a spin-$\frac{1}{2}$ charged particle is a one-loop prediction that is free of both ultraviolet (UV) divergences and infrared (IR) divergences, that is, it is a physical observable. However, it has been shown in~\cite{OP2} that LV induces contributions to the anomalous magnetic moments of leptons and quarks that are not free of IR divergences, showing that these quantities are no longer observable. We think that this type of result constitute a strong incentive to study one-loop LV effects on \textit{standard observables} (in the sense that they are invariant under both observer and particle Lorentz transformations) beyond the first order in the  $T^{\mu_1 \mu_2 \cdots}$ coefficients. In this work, we are interested in studying the two-point $A^a_\mu A^b_\nu$ vertex function in the context of pure Yang-Mills theories that incorporate all independent Lorentz violating interactions of renormalizable type. In particular, we will focus on the one-loop structure of the vacuum polarization tensor by considering all effects up to second order in the $T^{\mu_1 \mu_2 \cdots}$ coefficients of the theory. To carry out this program, we will adopt the background field method (BFM)~\cite{BFM}, which is a gauge-fixing procedure that allows us to maintain gauge invariance with respect to the background fields. This method greatly simplifies the renormalization program, since the renormalization constants associated with the gauge fields and the coupling constant of the gauge group are related to each other. Besides studying the one-loop renormalizability of the theory, we will put special attention on the impact of contributions proportional to $T^2=T^{\mu_1 \mu_2 \cdots}T_{\mu_1 \mu_2 \cdots}$ to the beta function associated with the coupling constant $g$ of the $SU(N)$ gauge group. To study the decoupling or non-decoupling nature of the new physics effects, the beta function will be analyzed from the perspective of both a mass-independent and a mass-dependent renormalization scheme. The one-loop renormalization of pure Yang-Mills theories has already been studied at first order in the Lorentz violating parameters in~\cite{YMER}. The one-loop renormalization of the QCD extension has been also studied by the same authors in~\cite{QCDER}.\\

The rest of the paper has been organized as follows. In Sec.~\ref{mYME}, the structure of the $SU(N)$-invariant Lagrangian that incorporates LV interactions of renormalizable dimension is studied. The Feynman rules in the BFM are derived. Secs.~\ref{Loop} and \ref{RBF} are devoted to study the one-loop structure of the theory. Finally, in Sec.~\ref{C} a summary is presented.

\section{The minimal Yang-Mills theory extension}
\label{mYME}The minimal Yang-Mills extension (mYME) is given by the following effective action:
\begin{equation}
\label{A}
S_{YME}[A^a_\mu]=\int d^4x{\cal L}_{YME}\, ,
\end{equation}
where
\begin{equation}
\label{L}
{\cal L}_{YME}={\cal L}_{YM}+{\cal L}^{CPT-Even}_{YMLV}+{\cal L}^{CPT-Odd}_{YMLV}+{\cal L}_{GF}+{\cal L}_{FPG}\, .
\end{equation}
In this expression,
\begin{eqnarray}
\label{LYM}
{\cal L}_{YM}&=&-\frac{1}{2}Tr[F_{\mu \nu}F^{\mu \nu}]\, , \\
\label{LYME}
{\cal L}^{CPT-Even}_{YMLV}&=&-\frac{1}{2}(k_F)^{\mu \nu \lambda \rho}Tr[F_{\mu \nu}F_{\lambda \rho}]\, ,\\
\label{LYMO}
{\cal L}^{CPT-Odd}_{YMLV}&=&-\frac{1}{2}(k_{AF})_{\kappa}\epsilon^{\kappa \lambda \mu \nu}Tr[A_\lambda F_{\mu \nu}+\frac{2}{3}ig A_\lambda A_\mu A_\nu]\, ,
\end{eqnarray}
 where $A_\mu=T^aA^a_\mu$ and $F_{\mu \nu}=T^aF^a_{\mu \nu}$. On the other hand, ${\cal L}_{GF}$ and ${\cal L}_{FPG}$ are the gauge-fixing Lagrangian and the Faddeev-Popov Lagrangian, respectively, which will be defined below. In addition, $F^a_{\mu \nu}$ are the Yang-Mills curvatures, which are given by:
\begin{equation}
F^a_{\mu \nu}=\partial_\mu A^a_\nu- \partial_\nu A^a_\mu +gf^{abc}A^b_\mu A^c_\nu\, .
\end{equation}
We will adopt the normalization $Tr[T^aT^b]=\frac{\delta^{ab}}{2}$. The $(k_F)^{\mu \nu \lambda \rho}$ tensor is dimensionless and has the same  symmetries as the Riemann tensor, so it satisfies the algebraic Bianchi identity
\begin{equation}
\label{BianchiI}
(k_F)_{\mu \nu \lambda \rho}+(k_F)_{\mu  \lambda \rho \nu}+(k_F)_{\mu \rho \nu \lambda}=0\, .
\end{equation}
For purposes of renormalization, it is convenient to work out with the $SO(1,3)$ irreducible parts of the Riemann-like tensor. So, in terms of its irreducible parts, $(k_F)_{\mu \nu \lambda \rho}$ can be decomposed as follows:
\begin{equation}
(k_F)_{\mu \nu \lambda \rho}=(\hat{k}_F)_{\mu \nu \lambda \rho}+(\tilde{k}_F)_{\mu \nu \lambda \rho}+\frac{1}{6}\left(g_{\mu \lambda} g_{\nu \rho}- g_{\mu \rho} g_{\nu \lambda} \right)\bar{k}_F\, ,
\end{equation}
where
\begin{equation}
(\tilde{k}_F)_{\mu \nu \lambda \rho}=\frac{1}{2}\left[g_{\nu \lambda}(k_F)_{\mu \rho}-g_{\nu \rho}(k_F)_{\lambda \mu}+
g_{\mu \rho}(k_F)_{\nu \lambda} -g_{\mu \lambda}(k_F)_{\nu \rho}\right]\, .
\end{equation}
In the above expressions, $(\hat{k}_F)_{\mu \nu \lambda \rho}$ is a Weyl-like tensor, which has the same symmetries as $(k_F)_{\mu \nu \lambda \rho}$, but it is defined so that every tensor contraction between indices gives zero. On the other hand, $(k_F)_{\nu \rho}=g^{\mu \lambda}(k_{F})_{\mu \nu \lambda \rho}$ is a symmetric tensor, analogous of the Ricci tensor. In addition, $\bar{k}_F=g^{\mu \nu}(k)_{\mu \nu}$ is the analogous of the scalar curvature. In terms of the irreducible parts of the $(k_F)_{\mu \nu \lambda \rho}$ tensor, the Lagrangian (\ref{LYME}) can be written as follows:
\begin{equation}
{\cal L}^{CPT-Even}_{YMLV}=-\frac{1}{2}(\hat{k}_F)^{\mu \nu \lambda \rho}Tr[F_{\mu \nu}F_{\lambda \rho}]-(k_F)^{\mu \nu}Tr[F_{\mu \lambda}F^\lambda _{\, \, \, \, \nu}]-\frac{\bar{k}_F}{6}Tr[F_{\mu \nu}F^{\mu \nu}]\, .
\end{equation}
Actually the $\bar{k}_F$ coefficient does not contribute, since it can be removed from the theory through the following redefinitions:
\begin{eqnarray}
A^a_\mu \to \Omega^{-\frac{1}{2}}A^a_\mu \, ,&& \, \, \, \, \, g\to \Omega^{\frac{1}{2}}g \, , \\
(\hat{k}_F)_{\mu \nu \lambda \rho}\to \Omega (\hat{k}_F)_{\mu \nu \lambda \rho} \, ,&& \, \, \, \, \, (k_F)_{\mu \nu}\to \Omega (k_F)_{\mu \nu}\, ,
\end{eqnarray}
where $\Omega=1+\frac{\bar{k}_F}{3}$. From now on, we will assume that $\bar{k}_F=0$. On the other hand, the coefficient $(k_{AF})_\kappa$ transforms as a 4-vector under observer Lorentz transformations, but it is invariant under particle Lorentz transformations. Since this coefficient has mass units, it can lead to important nondecoupling effects at the one-loop level. We will pay special attention to the consequences of this fact.

\subsection{Implementation of the Background Field Method}
Under an infinitesimal transformation, the gauge fields transform as
\begin{equation}
\label{gt}
\delta A^a_\mu={\cal D}^{ab}_\mu \alpha^b\, ,
\end{equation}
where ${\cal D}^{ab}_\mu=\delta^{ab}\partial_\mu-gf^{abc}A^c_\mu$ is the covariant derivative in the adjoint representation of $SU(N)$ and $\alpha^a$ are the gauge parameters. The BFM consists in decomposing the gauge fields into a classical part, $A^a_\mu$, and a quantum part, $Q^a_\mu$, $A^a_\mu \to A^a_\mu +Q^a_\mu$, so Eq.(\ref{gt}) becomes,
\begin{equation}
\delta (A^a_\mu+Q^a_\mu)=\left(\delta^{ab}\partial_\mu-gf^{abc}\left(A^c_\mu+Q^c_\mu\right)\right) \alpha^b
\end{equation}
or
\begin{eqnarray}
\label{gt1}
\delta A^a_\mu&=&{\cal D}^{ab}_\mu \alpha^b \, , \\
\label{gt2}
\delta Q^a_\mu&=&gf^{abc}Q^b_\mu \alpha^c\, ,
\end{eqnarray}
which shows that the $A^a_\mu$ fields transform as gauge fields, whereas the $Q^a_\mu$ fields transform as matter fields in the adjoint representation of $SU(N)$. On the other hand, the gauge curvatures becomes:
\begin{equation}
F^a_{\mu \nu} \to F^a_{\mu \nu}+{\cal D}^{ab}_\mu Q^b_\nu-{\cal D}^{ab}_\nu Q^b_\mu+gf^{abc}Q^b_\mu Q^c_\nu\, ,
\end{equation}
which transform in the adjoint representation of $SU(N)$. The $Q^a_\mu$ fields appear integrated in the fundamental path integral, so they are the quantum fields. On the other hand, the classical fields $A^a_\mu$ act as sources with respect to which the Green's functions of the theory are derived, that is, they represent the extern legs of such functions.\\

The BFM allows us to fix the gauge for the quantum fields $Q^a_\mu$ covariantly under the $SU(N)$ group. For this, we define the following gauge-fixing functions:
\begin{equation}
f^a={\cal D}^{ab}_\mu Q^{b \mu}\, ,
\end{equation}
which transform in the adjoint representation of $SU(N)$. In this way, the gauge-fixing Lagrangian
\begin{equation}
\label{gf}
{\cal L}_{GF}=-\frac{1}{2\xi}f^a f^a \, ,
\end{equation}
is invariant under the gauge transformations (\ref{gt1}). Here, $\xi$ is the gauge parameter.\\

In the BFM gauge, the Faddeev-Popov Lagrangian is given by:
\begin{equation}
\label{fpg}
{\cal L}_{FPG}=-\bar{c}^a{\cal D}^{ab}_\mu {\cal D}^{bc\mu} c^c+gf^{bcd}\bar{c}^a{\cal D}^{ab}_\mu Q^{c\mu}c^d\, ,
\end{equation}
where $c^a$ and $\bar{c}^a$ are the ghost and anti-ghost pairs of anticommuting fields, respectively.\\

\subsection{The renormalized Lagrangian and counterterm Lagrangian}

Since the quantum fields $Q^a_\mu$ only appear inside loops, the renormalization program is implemented on the background fields $A^a_\mu$. Also, at the one-loop level it is not necessary to introduce a renormalization for the ghost and anti-ghost fields, so we do not introduce a counterterm for the ghost sector.\\

Let $\left\{A^a_{B\mu}, g_B, (\hat{k}_{FB})^{\mu \nu \lambda \rho}, (k_{FB})^{\mu \nu}, (k_{AFB})_\kappa\right\}$ and $\left\{A^a_{\mu}, g, (\hat{k}_{F})^{\mu \nu \lambda \rho}, (k_{F})^{\mu \nu}, (k_{AF})_\kappa\right\}$ be the bare and renormalized gauge fields and coupling constants, respectively, which are related through the renormalization constants $\left\{Z_A, Z_g, Z^{\mu \nu \lambda \rho}_{F \alpha \beta \gamma \delta}, Z^{\mu \nu}_{F \alpha \beta}, Z^\kappa_{{AF} \alpha}\right\}$ as follows:
\begin{eqnarray}
A^a_{B\mu}&=&Z^{\frac{1}{2}}_A A^a_\mu\, , \\
 g_B&=&Z_g g \, ,
\end{eqnarray}
\begin{eqnarray}
(\hat{k}_{FB})^{\mu \nu \lambda \rho}&=&Z^{\mu \nu \lambda \rho}_{F \alpha \beta \gamma \delta}(\hat{k}_{F})^{\alpha \beta \gamma \delta}\, , \\
(k_{FB})^{\mu \nu}&=&Z^{\mu \nu}_{F \alpha \beta}(k_{F})^{\alpha \beta}\, , \\
(k_{AFB})_\kappa&=&Z^\alpha_{{AF} \kappa}(k_{AF})_\alpha\, .
\end{eqnarray}
One of the great advantages of BFM is the gauge invariance of the theory with respect to background fields $A^a_\mu$. As a consequence, we have the simple relation
\begin{equation}
F^a_{B\mu \nu}=Z^{\frac{1}{2}}_A F^a_{\mu \nu}\, ,
\end{equation}
which in turns implies that
\begin{equation}
\label{relation}
Z_g=Z^{-\frac{1}{2}}_A \,.
\end{equation}
So, the bare Lagrangian is given by
\begin{equation}
{\cal L}_{BYME}={\cal L}_{YME}+{\cal L}^{c.t.}_{YME}\, ,
\end{equation}
where ${\cal L}_{YME}$ is the renormalized Lagrangian given by Eq.(\ref{L}) and ${\cal L}^{c.t.}_{YME}$ is the counterterm Lagrangian, which is given by
\begin{eqnarray}
{\cal L}^{c.t.}_{YME}&=&-\frac{1}{2}\delta_ATr[F_{\mu \nu} F^{\mu \nu}]-\frac{1}{2}\left(\delta \hat{k}_F\right)^{\mu \lambda \nu \rho}Tr[F_{\mu \nu}F_{\lambda \rho}]
-\left(\delta k_F\right)^{\mu \nu}Tr[F_{\mu \lambda}F^{\lambda}_{\, \, \nu}]\nonumber \\
&&-\frac{1}{2}\left(\delta k_{AF}\right)_\kappa \epsilon^{\kappa \lambda \mu \nu}Tr[A_\lambda F_{\mu \nu}+\frac{2}{3}igA_\lambda A_\mu A_\nu]\, ,
\end{eqnarray}
where we have introduced the following definitions:
\begin{eqnarray}
\delta_A&=&Z_A-1\, , \\
(\delta\hat{k}_{F})_{\mu\lambda\nu\rho}&=& \left(Z_A Z^{\alpha \beta \gamma \delta}_{F\, \mu \lambda \nu \rho}-\delta^\alpha_\mu \delta^\beta_\lambda \delta^\gamma_\nu \delta^\delta_\rho \right)(\hat{k}_F)_{\alpha \beta \gamma \delta}\, ,\\
(\delta k_{F})_{\mu\nu}&=&\left(Z_A Z^{\alpha \beta}_{F\, \mu \nu}-\delta^\alpha_\mu \delta^\beta_\nu \right)(k_F)_{\alpha \beta}\, ,\\
(\delta k_{AF})_{\kappa}&=&\left(Z_A Z^\alpha_{AF\, \kappa}-\delta^\alpha_\kappa\right)(k_{AF})_\alpha \, .
\end{eqnarray}
Notice that the counterterm Lagrangian is gauge invariant. This fact implies, through Eq.(\ref{relation}), that the beta function associated with the coupling constant $g$ is determined by the renormalization constant $Z_A$. So it can be determined from a direct calculation of the vacuum polarization function. As commented in the introduction, we will extend this study to the complete two-point vertex function $A^a_\mu A^b_\nu$, which, up to second order in the Lorentz violating coefficients can be written as follows:
\begin{equation}
\Pi^{YME\, ab}_{\mu \nu}(q)=\Pi^{ab}_{\mu \nu}(q)+\Pi^{LV(1)\, ab}_{\mu \nu}(q)+\Pi^{LV(2)\, ab}_{\mu \nu}(q)\, ,
\end{equation}
where the first term represents the usual contribution, whereas the second and third terms include all effects of first- and second order in the Lorentz violating coefficients, respectively. We will see below that the second order term $\Pi^{LV(2)\, ab}_{\mu \nu}(q)$ contains a part that modifies the vacuum polarization and thus it introduces modifications in the usual beta function. The study of the consequences of these contributions is an important objective of this work.\\

We now proceed to derive the Feynman rules needed for the calculation on the one-loop correction to the two-point vertex function $\Pi^{YME\, ab}_{\mu \nu}(q)$. In the BFM gauge, the standard Yang-Mills sector is given by:
\begin{equation}
{\cal L}_{YM}=-\frac{1}{4}F^a_{\mu \nu}F^{a\mu \nu}-\frac{1}{2}F^a_{\mu \nu}{\cal Q}^{a\mu \nu}-\frac{g}{2}f^{abc}\left(F^a_{\mu \nu}+{\cal Q}^a_{\mu \nu}\right)Q^{b\mu}Q^{c\nu}-\frac{1}{4}{\cal Q}^a_{\mu \nu}{\cal Q}^{a\mu \nu}-\frac{g}{4}f^{abc}f^{ade}Q^b_\mu Q^c_\nu Q^{d\mu}Q^{e\nu}\, ,
\end{equation}
where ${\cal Q}^a_{\mu \nu}={\cal D}^{ab}_\mu Q^b_\nu-{\cal D}^{ab}_\nu Q^b_\mu$. On the other hand, the LV Lagrangians become:
\begin{eqnarray}
{\cal L}^{CPT-Even}_{YMLV}&=&-\frac{1}{4}(\hat{k}_F)^{\mu \nu \lambda \rho}\left(F^a_{\mu \nu}F^a_{\lambda \rho} +2gf^{abc}F^a_{\mu \nu}Q^b_\lambda Q^c_\rho+4\left({\cal D}^{ab}_\mu Q^b_\nu\right)\left({\cal D}^{ab}_\lambda Q^b_\rho\right)  \right)\nonumber \\
&&-\frac{1}{2}(k_F)^{\mu \nu}\left(F^a_{\mu \lambda}F^{\lambda}_{a\, \, \nu}+2gf^{abc}F^a_{\mu \lambda}Q^{b\lambda}Q^c_\nu+{\cal Q}^a_{\mu \lambda}{\cal Q}^{\lambda}_{a\, \, \nu}\right)+\cdots\, ,
\end{eqnarray}
where ellipsis indicate terms that do not contribute to the $\Pi^{YME\, ab}_{\mu \nu}(q)$ vertex function. In addition,
\begin{eqnarray}
{\cal L}^{CPT-Odd}_{YMLV}&=&-\frac{1}{4}(k_{AF})_\kappa \epsilon^{\kappa\lambda \mu \nu}\Big\{ A^a_{\lambda}F^a_{\mu \nu}-\frac{g}{3}f^{abc}A^a_\lambda A^b_\mu A^c_\nu+2(A^a_\lambda+Q^a_\lambda){\cal D}^{ab}_\mu Q^b_\nu \nonumber \\
&&+Q^a_\lambda F^a_{\mu \nu}-gf^{abc}\left(A^a_\lambda A^b_\mu Q^c_\nu+\frac{2}{3}Q^a_\lambda Q^b_\mu Q^c_\nu\right)\Big\}\, .
\end{eqnarray}

The vertices that contribute to the two-point $\Pi^{YME\, ab}_{\mu \nu}(q)$ vertex function up to second order in the LV coefficients are: $QQ$, $AQQ$, and $AAQQ$, whose corresponding Feynman rules are shown in Fig.\ref{RFR}. The Lorentz tensors that appear in these Feynman rules are given by:

\begin{figure}[ht]
\center
\includegraphics[trim= 20mm -10mm 50mm 0mm, scale=0.65,clip, angle=270]{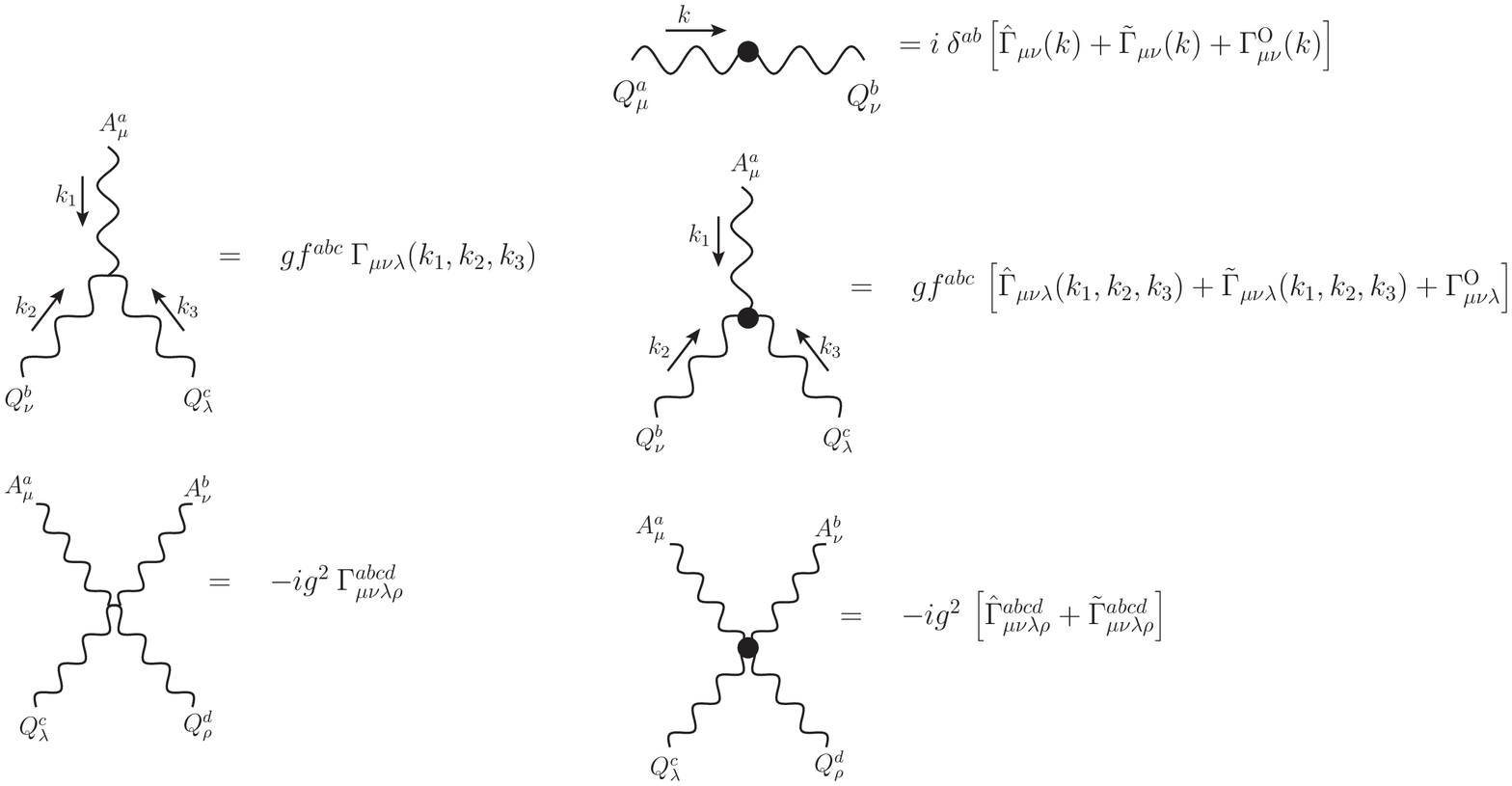}
\caption{\label{RFR} {\footnotesize Feynman rules needed to calculate the one-loop amplitude of the $A^a_\mu A^b_\nu$ vertex fuction in the BFM-gauge.}}
\end{figure}

\begin{eqnarray}\label{AQQYMBFM}
		\Gamma_{\mu\nu\lambda}(k_{1},k_{2},k_{3})&=& g_{\lambda \mu} \Big(k_{3}-k_{1}+\frac{k_{2}}{\xi
		}\Big)_{\nu}+g_{\mu \nu } \Big(k_{1}-k_{2}-\frac{k_{3}}{\xi }\Big)_{\lambda}+g_{\lambda \nu } (k_{2}-k_{3})_{\mu} \: ,
	\end{eqnarray}	
	\begin{eqnarray}\label{AAQQYMBFM}
		\Gamma^{abcd}_{\mu\nu\lambda\rho}&=&f^{ace}f^{bde}\Big(g_{\mu\nu}g_{\lambda\rho} -g_{\mu\rho}g_{\nu\lambda}+\frac{1}{\xi}g_{\mu\lambda}g_{\nu\rho}\Big)+f^{ade}f^{bce}\Big(g_{\mu\nu}g_{\lambda\rho} -g_{\mu\lambda}g_{\nu\rho}+\frac{1}{\xi}g_{\mu\rho}g_{\nu\lambda}\Big)\nonumber \\
&&+f^{abe}f^{cde}(g_{\mu\lambda}g_{\nu\rho} -g_{\mu\rho}g_{\nu\lambda})\: ,
	\end{eqnarray}
\begin{eqnarray}
		\hat{\Gamma}_{\mu\nu}(k) &=& -2 (\hat{k}_{F})_{\mu\lambda\nu\rho}k^{\lambda}k^{\rho}\: , \label{GamaQQHat} \\
		\tilde{\Gamma}_{\mu\nu}(k)&=&\left(k^2\delta^\lambda_\mu \delta^\rho_\nu +g_{\mu \nu}k^\lambda k^\rho -k_\mu k^\lambda \delta^\rho_\nu
		-k_\nu k^\lambda \delta^\rho_\mu  \right)(k_F)_{\lambda \rho}\: \: \: \label{GammaQQTilde}\\
		\Gamma_{\mu \nu}^{\mathrm{O}}(k)&=&-i\: (k_{AF})^{\kappa}\epsilon_{\kappa\mu\rho\nu}k^{\rho},\label{OQQ}
\end{eqnarray}
	
		\begin{eqnarray}
		\hat{\Gamma}_{\mu\nu\lambda}(k_{1},k_{2},k_{3})&=&-2\big[k_{1} ^{\rho}(\hat{k}_{F})_{\rho\mu\nu\lambda}+k_{2}^{\rho}(\hat{k}_{F})_{\rho\nu\lambda\mu}+k_{3}^{\rho}(\hat{k}_{F})_{\rho\lambda\mu\nu}\big]\: , \label{GammaAQQHat}\\
		\tilde{\Gamma}_{\mu\nu\lambda}(k_{1},k_{2},k_{3})&=& \:\:\: \big[(k_{F})_{\mu\nu}g_{\rho\lambda}+g_{\mu\nu}(k_{F})_{\rho\lambda}\big](k_{2}-k_{1})^{\rho}+\big[(k_{F})_{\mu\lambda}g_{\rho\nu}+g_{\mu\lambda}(k_{F})_{\rho\nu}\big](k_{1}-k_{3})^{\rho} \nonumber \\
		&&+\big[(k_{F})_{\nu\lambda}g_{\rho\mu}+g_{\nu\lambda}(k_{F})_{\rho\mu}\big](k_{3}-k_{2})^{\rho}\: , \label{GammaAQQTilde} \\
		\Gamma^{\mathrm{O}}_{\mu\nu\lambda} &=&-i\:  (k_{AF})^{\kappa}\epsilon_{\kappa\mu\nu\lambda}  \: \: \: , \label{OAQQ}
	\end{eqnarray}
\begin{eqnarray}
		\hat{\Gamma}^{abcd}_{\mu\nu\lambda\rho}&=&2[f^{abe}f^{cde}(\hat{k}_{F})_{\mu\nu\lambda\rho} +f^{ace}f^{bde}(\hat{k}_{F})_{\mu\lambda\nu\rho}+f^{ade}f^{bce}(\hat{k}_{F})_{\mu\rho\nu\lambda} ]\: ,\label{GammaAAQQHat}\\
	\tilde{\Gamma}^{abcd}_{\mu\nu\lambda\rho}&=&2[f^{abe}f^{cde}(\tilde{k}_{F})_{\mu\nu\lambda\rho} +f^{ace}f^{bde}(\tilde{k}_{F})_{\mu\lambda\nu\rho}+f^{ade}f^{bce}(\tilde{k}_{F})_{\mu\rho\nu\lambda} ]\: .\label{GammaAAQQTilde}
\end{eqnarray}

\begin{figure}[h]
	\includegraphics[trim= -12mm 248mm 0mm 30mm, scale=0.70,clip, angle=0]{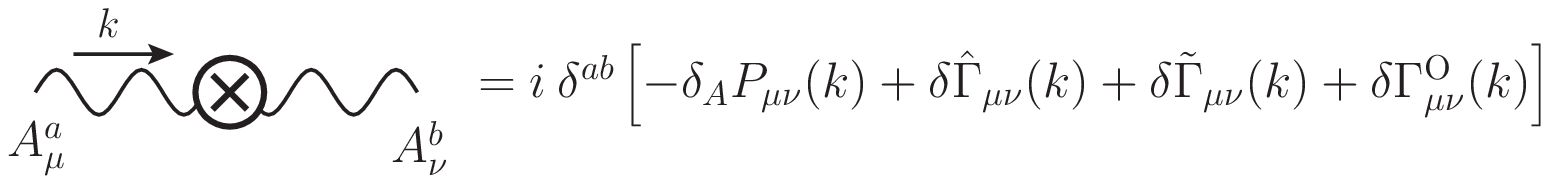}
	\caption{\small Feynman rule of the counterterm associated with the $A^{a}_{\mu}A^{b}_{\nu}$ coupling.}
	\label{CFR}
\end{figure}

On the other hand, the Feynman rule associated with the counterterm is shown in Fig.\ref{CFR}.

The various quantities appearing in Fig.\ref{CFR} are given by:
\begin{eqnarray}
	P_{\mu\nu}(k)&=&\quad k^{2}g^{\mu\nu}-k^{\mu}k^{\nu}\: ,  \label{Pmn} \\
    \delta\hat{\Gamma}_{\mu\nu}(k) &=& -2 (\delta\hat{k}_{F})_{\mu\lambda\nu\rho}k^{\lambda}k^{\rho}\: , \label{GamaQQHatct} \\
    \delta\tilde{\Gamma}_{\mu\nu}(k)&=&\left(k^2\delta^\lambda_\mu \delta^\rho_\nu +g_{\mu \nu}k^\lambda k^\rho -k_\mu k^\lambda \delta^\rho_\nu
    -k_\nu k^\lambda \delta^\rho_\mu  \right)(\delta k_F)_{\lambda \rho}\: , \label{GamaQQTildect}\\
    \delta\Gamma^{\mathrm{O}}_{\mu\nu}(k)&=&\: -i\: (\delta k_{AF})^{\kappa}\epsilon_{\kappa\mu\rho\nu}k^{\rho}\: \: \:  . \label{GammaQQoddct}
\end{eqnarray}
 In Table \ref{TFV}, we present a summary of the Feynman rules used in the calculation of the $A^{a}_{\mu}A^{b}_{\nu}$ vertex function.

	\renewcommand{\arraystretch}{1.7}
\begin{table}[h]
	\begin{center}
		\begin{tabular}{| c|   c| c |c|}
			\hline
			Vertex Function &  $\mathcal{L}_{\mathrm{YM}}$  & $\mathcal{L}^{\mathrm{CPT-Even}}_{\mathrm{YM-LV}}$ &$\mathcal{L}^{\mathrm{CPT-Odd}}_{\mathrm{YM-LV}}$ \\
			
			\hline
			
		 \footnotesize $Q^{a}_{\mu}(k)Q^{b}_{\nu}(k)$ &  ---------  & \footnotesize  $i\delta^{ab}\left(\hat{\Gamma}_{\mu\nu}(k)+\tilde{\Gamma}_{\mu\nu}(k)\right)$ , Eqs.(\ref{GamaQQHat},\ref{GammaQQTilde})  &\footnotesize  $i\delta^{ab}\Gamma^{\mathrm{O}}_{\mu\nu}(k)\:$,  Eq.(\ref{OQQ})  \\
			
			\hline
			
		 \footnotesize $A^{a}_{\mu}(k_{1})Q^{b}_{\nu}(k_{2})Q^{c}_{\lambda}(k_{3})$& \footnotesize  $g f^{abc} \Gamma_{\mu\nu\lambda}(k_{1},k_{2},k_{3})$ Eq. (\ref{AQQYMBFM}) & \footnotesize  $gf^{abc}\left(\hat{\Gamma}_{\mu\nu\lambda}(k_{1},k_{2},k_{3})+\tilde{\Gamma}_{\mu\nu\lambda}(k_{1},k_{2},k_{3})\right)$, Eqs.(\ref{GammaAQQHat},\ref{GammaAQQTilde})  & \footnotesize $g f^{abc}\Gamma^{\mathrm{O}}_{\mu\nu\lambda} $,  Eq.(\ref{OAQQ})  \\
			
			\hline
			
		 \footnotesize  $A^{a}_{\mu}A^{b}_{\nu}Q^{c}_{\lambda}Q^{d}_{\rho}$&\footnotesize  $-i g^{2} \Gamma^{abcd}_{\mu\nu\lambda\rho}$,  Eq.(\ref{AAQQYMBFM})&\footnotesize  $-i g^{2}\left( \hat{\Gamma}^{abcd}_{\mu\nu\lambda\rho}+\tilde{\Gamma}^{abcd}_{\mu\nu\lambda\rho}\right)$, Eqs.(\ref{GammaAAQQHat},\ref{GammaAAQQTilde})  &  ---------   \\
		
			\hline
			
			 \footnotesize	\textbf{C.t.}	$A^{a}_{\mu}(k)A^{b}_{\nu}(k)$& \footnotesize $-i\: \delta^{ab} \delta_{A}P_{\mu\nu}(k)$ Eq. (\ref{Pmn}) & \footnotesize $i\: \delta^{ab}\left( \delta\hat{\Gamma}_{\mu\nu}+\delta\tilde{\Gamma}_{\mu\nu}\right)$,  Eqs.(\ref{GamaQQHatct},\ref{GamaQQTildect})&\footnotesize  $ i \delta^{ab}\delta\Gamma^{\mathrm{O}}_{\mu\nu}(k)$,  Eq.(\ref{GammaQQoddct})  \\
			\hline
		\end{tabular}
		\caption{Vertex functions contributing to the $A^{a}_{\mu}A^{b}_{\nu}$ coupling at the one-loop level.} \label{TFV}
	\end{center}
\end{table}

\section{The one-loop $\Pi^{YME\, ab}_{\mu \nu}(q)$ vertex function}
\label{Loop}
\begin{figure}[h]
	\includegraphics[trim= 20mm 150mm 0mm 20mm, scale=0.70,clip, angle=0]{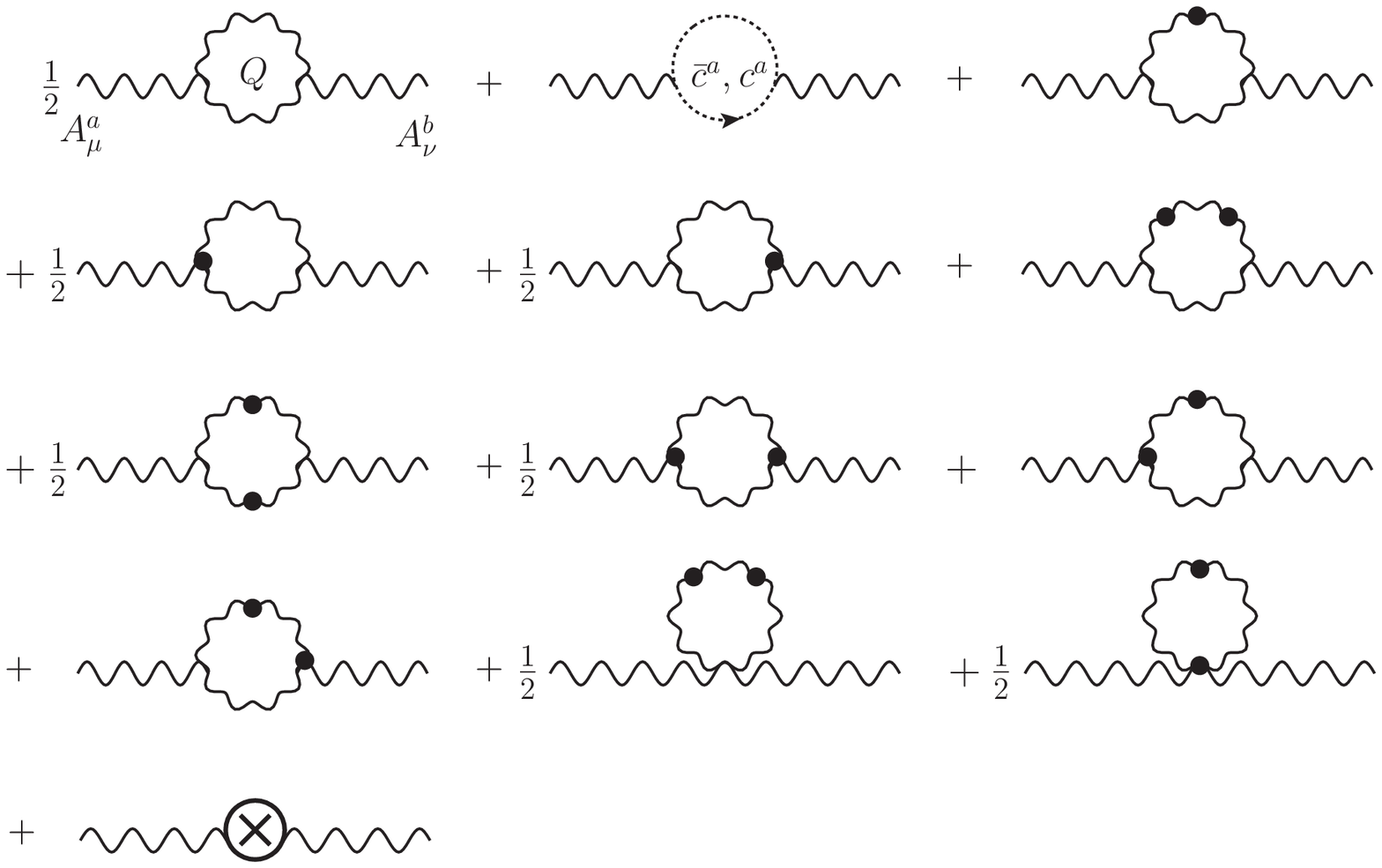}
	\caption{ Feynman diagrams contributing to the $A^a_\mu A^b_\nu$ vertex function at one-loop level up to second order in the Lorentz coefficients. The first two diagrams represent the usual contribution, whereas those diagrams with one black dot and two black dots represent contributions of first and second order in the Lorentz coefficients, respectively. The contribution of the counterterm has been included.}
	\label{AA012Order}
\end{figure}

In this Section, we study the one-loop quantum fluctuations induced by LV effects on the $A^a_\mu A^b_\nu$ vertex function. Effects of up to second order in the Lorentz violating coefficients will be considered. Up to second order in the Lorentz coefficients, the contribution to this two-point vertex function is given by the Feynman diagrams shown in Fig.~\ref{AA012Order}. The structure of the corresponding amplitude is dictated by $SU(N)$-gauge invariance and can be written as follows:
\begin{eqnarray}
\Pi^{YME\, ab}_{\mu \nu}(q,\xi)&=&i\delta^{ab}\left[\Pi_{\mu \nu}(q,\xi)+\Pi^{E(1)}_{\mu \nu}(q,\xi)+\Pi^{E(2)}_{\mu \nu}(q,\xi)+\Pi^{O(1)}_{\mu \nu}(q,\xi)+\Pi^{O(2)}_{\mu \nu}(q,\xi)+\Pi^{EO(2)}_{\mu \nu}(q,\xi)\right]\nonumber \\
&&+i\delta^{ab}\left[-\delta_A P_{\mu \nu}(q)+\delta \hat{\Gamma}_{\mu \nu}+\delta \tilde{\Gamma}_{\mu \nu}+\delta \Gamma^O_{\mu \nu}\right]\, .
\end{eqnarray}
Notice that we have included the contribution of the counterterms. In these expressions, the label $E(O)$ stands for CPT-Even (CPT-Odd) contribution. Also, the $(1)$ and  $(2)$ labels indicate contributions of first- and second order in the Lorentz coefficients, respectively. Notice that interference effects between CPT-Even and CPT-Odd terms, which can be generated to second order in the LV coefficients, are also considered. We now proceed to describe each of these contributions. For comparison purposes, and also by clarity, the usual contribution, as well as the first order contribution already studied in Ref.~\cite{QCDER}, will be presented. In all cases, exact expressions calculated in the general BFM-gauge will be presented. We have performed our calculations using the FeynCalc package~\cite{FCP}.

\subsection{The usual contribution $\Pi_{\mu \nu}(q,\xi)$}
In the general BFM-gauge, the one-loop contribution of the usual theory is given through the first two diagrams shown in Fig.~\ref{AA012Order}. The vacuum polarization tensor function is given by:
\begin{equation}
\Pi_{\mu \nu}(q,\xi)=\Pi(q^2,\xi)P_{\mu \nu}(q)\, ,
\end{equation}
where $\Pi(q^2,\xi)$ is the vacuum polarization function, which is given by:
\begin{equation}
\label{Pi}
\Pi(q^2,\xi)=\frac{g^2}{(4\pi)^2}C_2(G)\left[\frac{11}{3}\Delta -\frac{11}{3}\log\left(-\frac{q^2}{\hat{\mu}^2}\right)+\frac{67}{9}-\frac{1}{4}\left(1-\xi\right)\left(7+\xi\right)\right]\, .
\end{equation}
In this expression,
\begin{equation}
\label{Del}
\Delta=\frac{1}{\epsilon}-\gamma_E+\log(4\pi)\, ,
\end{equation}
with $\gamma_E$ the Euler-Mascheroni constant and $\hat{\mu}$ the scale of dimensional regularization. The quantity $C_2(G)$ has to do with the normalization of $SU(N)$ generators in the adjoint representation through the relation $f^{acd}f^{bcd}=C_2(G)\delta^{ab}$.

\subsection{First order contributions}
At first order in the LV coefficients, the contributions are given by those diagrams characterized by one black point shown in Fig~\ref{AA012Order}.

\subsubsection{CPT-Even contribution}
The CPT-Even contribution can be written as follows:
\begin{equation}
\Pi^{E(1)}_{\mu \nu}(q,\xi)=f_1(q^2,\xi)\hat{\Gamma}_{\mu \nu}(q)+f_2(q^2,\xi)\tilde{\Gamma}_{\mu \nu}(q)+\Pi^{E(1)}(q^2,\xi)P_{\mu \nu} (q) \, ,
\end{equation}
where
\begin{equation}
\label{PiE1}
\Pi^{E(1)}(q^2,\xi)=f_3(\xi)(k_F)_{\alpha \beta}\left(\frac{q^\alpha q^\beta}{q^2}\right)\, ,
\end{equation}
 and the $\hat{\Gamma}_{\mu \nu}$ and $\tilde{\Gamma}_{\mu \nu}$ tensors are given in Eqs.(\ref{GamaQQHat}) and (\ref{GammaQQTilde}), respectively. The form factors $f_1(q^2,\xi)$ and $f_2(q^2,\xi)$ are UV divergent. On the other hand, the $f_3(\xi)$ form factor is free of UV divergences, which is a consequence of the fact that it is associated with a dimension-six interaction. This finite contribution, proportional to the usual vacuum polarization tensor structure $P_{\mu\nu}$, is associated with following dimension-six interaction:
\begin{equation}
	(k_F)^{\alpha \beta}{\cal D}^{ab}_\alpha F^b_{\lambda \rho} {\cal D}^{ac}_\beta F^{c\lambda \rho}\, .
\end{equation}
The $f_i(q^2,\xi)$ form factors are given in Appendix~\ref{AP}. It is worth mentioning that the term proportional to the Ricci tensor is not considered in Ref.~\cite{YMER}, since the authors take it equal to zero from the beginning. However, this is valid only at first order in the LV coefficients, but not to higher orders. Next, we will show that the Ricci structure is generated to second order by a specific indices contraction between two Weyl-like tensors, so  renormalization theory requires the presence of said coefficient in the classical Lagrangian if contributions beyond the first order are considered.\\

\subsubsection{CPT-Odd contribution}
On the other hand, the CPT-Odd contribution is given by:
\begin{equation}
\label{O(1)}
\Pi^{O(1)}_{\mu \nu}(q,\xi)=-\frac{ig^2C_2(G)}{2(4\pi)^2}\left[13-\xi(\xi+4)\right](k_{AF})^\lambda \epsilon_{\lambda \mu \rho \nu}q^\rho \, ,
\end{equation}
which is free of UV divergences. However, this result seems to be unique to the BFM-gauge, since it is divergent in the linear $R_\xi$-gauge, as shown in Ref.~\cite{YMER}. Although the result~(\ref{O(1)}) is free of UV divergences, the renormalization factor associated with the Lorentz coefficient $(k_{AF})_\kappa$ is UV divergent, which acquires a pole through the renormalization factor $Z_A$ of the gauge fields. This fact has already been pointed out in Ref.~\cite{QCDER}, where the one-loop structure of the QCD extension was studied to first order using the BFM-gauge.\\

Notice that the $\epsilon_{\lambda \mu \rho \nu}q^\rho$ tensor is symmetric in the pair of indices $\mu$ and $\nu$, as the interchange $\mu \leftrightarrow \nu$ must be accompanied by the change $q \rightarrow-q$.

\subsection{Second order contributions}
The second order contribution in the Lorentz coefficients is given by those diagrams in Fig.~\ref{AA012Order} characterized by two black dots.

\subsubsection{CPT-Even contribution}
The CPT-even contribution can be written as follows:
\begin{equation}
\label{SO}
\Pi^{E(2)}_{\mu \nu}(q,\xi)=\hat{\Gamma}^{(2)}_{\mu \nu}+\tilde{\Gamma}^{(2)}_{\mu \nu}+\left(\frac{q^\lambda q^\rho}{q^2}\right)\tilde{\Gamma}^{(2)}_{\mu \nu \lambda \rho}+\left(\frac{q^\lambda q^\rho q^\sigma q^\tau}{(q^2)^2}\right)\tilde{\Gamma}^{(2)}_{\mu \nu \lambda \rho \sigma \tau}+\Pi^{E(2)}(q^2,\xi)P_{\mu \nu}\, ,
\end{equation}
where
\begin{equation}
\label{PiE2}
\Pi^{E(2)}(q^2,\xi)=\Pi^{E(2)}_{Div}+\left(\frac{q^\alpha q^\beta}{q^2}\right)(k_1)_{\alpha \beta}+\left(\frac{q^\alpha q^\beta q^\sigma q^\tau}{(q^2)^2}\right)(k_2)_{\alpha \beta \sigma \tau}\, .
\end{equation}
In the above expressions,
\begin{eqnarray}
\label{Gamma2hat}
\hat{\Gamma}^{(2)}_{\mu \nu}&=&-2q^\lambda q^\rho (\hat{k}^{(2)}_F)_{\mu \lambda \nu \rho}\, , \\
\label{Gamma2tilde}
\tilde{\Gamma}^{(2)}_{\mu \nu}&=&\left(q^2\delta^\alpha_\mu \delta^\beta_\nu +g_{\mu \nu}q^\alpha q^\beta -q_\mu q^\alpha \delta^\beta_\nu
-q_\nu q^\alpha \delta^\beta_\mu  \right)(\tilde{k}^{(2)}_F)_{\alpha \beta}\, ,\\
\label{Gamma2tilde4}
\tilde{\Gamma}^{(2)}_{\mu \nu \lambda \rho}&=&\left(q^2\delta^\alpha_\mu \delta^\beta_\nu +g_{\mu \nu}q^\alpha q^\beta -q_\mu q^\alpha \delta^\beta_\nu
-q_\nu q^\alpha \delta^\beta_\mu  \right)(\tilde{k}^{(2)}_F)_{\alpha \beta \lambda \rho}\, , \\
\label{Gamma2tilde6}
\tilde{\Gamma}^{(2)}_{\mu \nu \lambda \rho \sigma \tau}&=&\left(q^2\delta^\alpha_\mu \delta^\beta_\nu +g_{\mu \nu}q^\alpha q^\beta -q_\mu q^\alpha \delta^\beta_\nu
-q_\nu q^\alpha \delta^\beta_\mu  \right)(\tilde{k}^{(2)}_F)_{\alpha \beta \lambda \rho \sigma \tau}\, ,
\end{eqnarray}
where the $(\hat{k}^{(2)}_F)_{\mu \lambda \nu \rho}$, $(\tilde{k}^{(2)}_F)_{\alpha \beta}$, $(\tilde{k}^{(2)}_F)_{\alpha \beta \lambda \rho}$, and $(\tilde{k}^{(2)}_F)_{\alpha \beta \lambda \rho \sigma \tau}$ tensors are made of contractions between the tree-level Weyl-like $(\hat{k}_F)_{\mu \nu \alpha \beta}$ tensor and the tree-level Ricci-like $(k_F)_{\mu \nu}$ tensor. The first of these tensors,  $(\hat{k}^{(2)}_F)_{\mu \lambda \nu \rho}$, is a Wely-like tensor, which emerges from the irreducible pieces of the following Riemann-like tensor:
\begin{equation}
(k^{(2)}_F)_{\mu \lambda \nu \rho}=l_1(q^2,\xi)(k^{(2)}_W)_{\mu \lambda \nu \rho}+l_2(q^2,\xi)(k^{(2)}_R)_{\mu \lambda \nu \rho}\, ,
\end{equation}
where $(k^{(2)}_W)_{\mu \lambda \nu \rho}$ and $(k^{(2)}_R)_{\mu \lambda \nu \rho}$ are Riemann-like tensors, which are given by:
\begin{eqnarray}
(k^{(2)}_W)_{\mu \lambda \nu \rho}&=&(\hat{k}_F)_{\mu \lambda \sigma \tau}(\hat{k}_F)_{\nu \rho}^{\, \, \, \, \, \,  \sigma \tau}+\frac{1}{2}
(\hat{k}_F)_{\mu \rho \sigma \tau}(\hat{k}_F)_{\nu \lambda}^{\, \, \, \, \, \,  \sigma \tau}-\frac{1}{2}
(\hat{k}_F)_{\mu \nu \sigma \tau}(\hat{k}_F)_{ \lambda \rho}^{\, \, \, \, \, \,  \sigma \tau}\, ,\\
(k^{(2)}_R)_{\mu \lambda \nu \rho}&=&(k_F)_{\mu \rho}(k_F)_{\nu \lambda}-(k_F)_{\mu \nu}(k_F)_{\lambda \rho}\, .
\end{eqnarray}
It can be shown that these tensors satisfy all symmetries of a Riemann-like tensor, the algebraic Bianchi identity (\ref{BianchiI}), and have non-zero double trace. The decomposition of $(k^{(2)}_F)_{\mu \lambda \nu \rho}$ into its irreducibles parts, leads to the following Weyl-like tensor
\begin{equation}
\label{Weyl2}
(\hat{k}^{(2)}_F)_{\mu \lambda \nu \rho}=l_1(q^2,\xi)(\hat{k}^{(2)}_W)_{\mu \lambda \nu \rho}+l_2(q^2,\xi)(\hat{k}^{(2)}_R)_{\mu \lambda \nu \rho}\, ,
\end{equation}
where $(\hat{k}^{(2)}_{W,R})_{\mu \lambda \nu \rho}$ are the Weyl-like parts of the $(k^{(2)}_{W,R})_{\mu \lambda \nu \rho}$ Riemann-like tensors. The Ricci-like part of $(k^{(2)}_F)_{\mu \lambda \nu \rho}$ is given by:
\begin{equation}
({k}^{(2)}_F)_{\mu \nu}=l_1(q^2,\xi)({k}^{(2)}_{W})_{\mu \nu}+l_2(q^2,\xi)({k}^{(2)}_{R})_{\mu \nu}\, ,
\end{equation}
where $({k}^{(2)}_{W,R})_{\mu \nu}$ are the Ricci-like parts of the $(k^{(2)}_{W,R})_{\mu \lambda \nu \rho}$ Riemann-like tensor. This Ricci-like contribution is considered inside of the $(\tilde{k}_F^{(2)})_{\alpha \beta}$ tensor appearing in the $\tilde{\Gamma}^{(2)}_{\mu \nu}$ tensor of Eq.~(\ref{Gamma2tilde}). As regards the analogous of scalar curvature, $\bar{k}^{(2)}_F$, its contribution is included in the $\Pi^{E(2)}$ scalar function that appears in Eq.~(\ref{SO}). The loop functions $l_1(q^2,\xi)$ and  $l_2(q^2,\xi)$  that appear in the above expressions are UV divergent. They are presented in Appendix~\ref{AP}.\\

Some comments are in order here. The Riemann-like tensor $(k^{(2)}_F)_{\mu \lambda \nu \rho}$ is the only source of a Weyl-like contribution, namely, the one given by Eq.~(\ref{Gamma2hat}), but it is not the only source of Ricci contributions and scalar curvature contributions. There are additional contributions to the Ricci and $P_{\mu \nu}$ Lorentz structures arising from sources different to the Riemann-like tensor $(k^{(2)}_F)_{\mu \lambda \nu \rho}$. Once all the Ricci-type contributions are added, we obtain
\begin{equation}
\label{Ricci2}
(\tilde{k}_F^{(2)})_{\mu \nu}=g_1(q^2,\xi)( \hat{k}_F)_{\mu \lambda \rho \sigma}( \hat{k}_F)_{\nu}^{\, \, \, \lambda \rho \sigma}+
g_2(q^2,\xi)( k_F)_{\mu \lambda}( k_F)_{\nu}^{\, \, \, \lambda}+
g_3(q^2,\xi)( \hat{k}_F)_{\mu \lambda \nu \sigma}( k_F)^{\lambda \sigma}
\end{equation}
being the $(\tilde{k}_F^{(2)})_{\mu \nu}$ tensor the one appearing in Eq.~(\ref{Gamma2tilde}). It is important to note the presence in Eq.~(\ref{Ricci2}) of a Ricci-like factor, namely, the one with form factor $g_1(q^2, \xi)$, which is independent of the tree-level $(k_F)_{\mu \nu}$ Ricci-like tensor. Since $g_1(q^2,\xi)$ is UV divergent, we cannot take the tree-level Ricci-like tensor $(k_F)_{\mu \nu}$ equal to zero, as assumed in Ref.~\cite{YMER}, if second order effects are considered. The three form factors $g_i(q^2,\xi)$ are UV divergent. They are given in Appendix~\ref{AP}.\\

On the other hand, the $\Pi^{E(2)}_{Div}$ form factor is given by:
\begin{equation}
\label{VPF2}
\Pi^{E(2)}_{Div}(q^2,\xi)=h_1(q^2,\xi)(\hat{k}_F)^2+h_2(q^2,\xi)(k_F)^2\, ,
\end{equation}
where
\begin{eqnarray}
(\hat{k}_F)^2&=&(\hat{k}_F)_{\alpha \beta \lambda \rho}(\hat{k}_F)^{\alpha \beta \lambda \rho}\, ,\\
(k_F)^2&=&(k_F)_{\alpha \beta}(k_F)^{\alpha \beta}\, .
\end{eqnarray}
The UV divergent loop functions $h_i(q^2,\xi)$ are given in Appendix~\ref{AP}.\\

As far as the 4-tensor and 6-tensor structures given in Eqs.~(\ref{Gamma2tilde4}) and (\ref{Gamma2tilde6}) are concerned, they are given by:
\begin{eqnarray}
	(\tilde{k}^{(2)}_F)_{\alpha \beta \lambda \rho}&=&\,\,\, g_4(\xi)\left[(\hat{k}_F)_{\alpha \sigma \lambda \tau}(\hat{k}_F)^{\, \, \, \, \sigma \, \, \, \tau}_{\beta \, \, \, \rho}+(\hat{k}_F)_{\beta\sigma \lambda \tau}(\hat{k}_F)^{\, \, \, \, \sigma \, \, \, \tau}_{\alpha \, \, \, \rho}\right]+g_5(\xi)(k_F)_{\alpha \beta}(k_F)_{\lambda \rho}\nonumber \\
	&&+g_6(\xi)\left[(\hat{k}_F)_{\alpha \lambda \rho \sigma}(k_F)_\beta ^{\, \, \, \sigma}+(\hat{k}_F)_{\beta \lambda \rho \sigma}(k_F)_\alpha^{\, \, \, \sigma}\right]+g_7(\xi)\left[(\hat{k}_F)_{\alpha \rho \beta \sigma}+(\hat{k}_F)_{\beta \rho \alpha \sigma}\right](k_F)_\lambda ^{\, \, \, \sigma}\, , \label{Ricci6} \\
	(\tilde{k}^{(2)}_F)_{\alpha \beta \lambda \rho \sigma \tau}&=&g_8(\xi)(\hat{k}_F)_{\alpha \sigma \lambda \omega}(\hat{k}_F)_{\beta \tau \rho}^{\, \, \, \, \, \, \, \, \, \, \, \omega}+g_9(\xi)(\hat{k}_F)_{\alpha \lambda \beta \rho}(k_F)_{\sigma \tau}+(\alpha \leftrightarrow \beta)\, .\label{Ricci8}
\end{eqnarray}
These observer-Lorentz structures characterize dimension-six and dimension-eight interactions, which are Ricci-like tensors with respect to the first two indices. The dimension-six interactions have the following structure:
\begin{eqnarray}
	&(\hat{k}_{F})^{\mu\sigma\alpha\tau}(\hat{k}_{F})_{\nu\sigma\beta\tau}{\cal D}^{ab}_{\alpha}F^b_{\mu\lambda}{\cal D}^{ac\beta}F^{c\:\lambda\nu}& \,  , \, \, 	(k_{F})^{\mu\nu}(k_{F})^{\alpha\beta}{\cal D}^{ab}_{\alpha}F^b_{\mu\lambda}{\cal D}^{ac}_{\beta}F^{c\:\lambda}{}_{\nu}\, \,\, , \, \, \nonumber\\	&(\hat{k}_{F})^{\mu\alpha\beta\sigma}(k_{F})_{\nu\sigma}{\cal D}^{ab}_{\alpha}F^b_{\mu\lambda}{\cal D}^{ac}_{\beta}F^{c\:\lambda\nu}& \, , \, \, 	(\hat{k}_{F})^{\mu\beta\nu\sigma}(k_{F})_{\alpha\sigma}{\cal D}^{ab\alpha}F^b_{\mu\lambda}{\cal D}^{ac}_{\beta}F^{c\:\lambda}{}_{\nu} \, ,
\end{eqnarray}
while those of dimension-eight have the form:
\begin{equation}
	(\hat{k}_{F})^{\mu\sigma\alpha\omega}(\hat{k}_{F})_{\nu\tau\beta\omega}{\cal D}^{ab}_{\alpha}{\cal D}^{bc}_{\sigma}F_{\mu\lambda}^{c}{\cal D}^{ad\beta}{\cal D}^{de\tau}F^{e\lambda\nu} \, , \, \, (\hat{k}_{F})^{\mu\alpha\nu\beta}(k_{F})^{\sigma\tau}{\cal D}^{ab}_{\alpha}{\cal D}^{bc}_{\sigma}F_{\mu\lambda}^{c}{\cal D}^{ad}_{\beta}{\cal D}^{de}_{\tau}F^{e\lambda}{}_{\nu}\, .
\end{equation}
The loop functions $g_i(\xi)$ $(i=4,\cdots, 9)$ are all free of UV divergences, which is in accordance with renormalization theory, as they characterize interactions of dimension higher than four. These functions, which depend on only the gauge parameter, are given in Appendix~\ref{AP}.\\

Finally, there are also dimension-six and dimension-eight interactions which contribute to the usual $P_{\mu \nu}$ tensor. These contributions emerge through the following tensors [see Eqs.~(\ref{SO},\ref{PiE2})]:
\begin{eqnarray}
\label{k1}
(k_1)_{\alpha \beta}&=&s_1(\xi)(\hat{k}_F)_{\alpha \lambda \rho \sigma}(\hat{k}_F)_\beta^{\, \, \, \, \lambda \rho \sigma}+s_2(\xi)(k_F)_{\alpha \lambda}(k_F)_\beta^{\, \, \, \, \lambda}+s_3(\xi)(\hat{k}_F)_{\alpha \lambda \beta \sigma}(k_F)^{\lambda \sigma}\, ,\\
\label{k2}
(k_2)_{\alpha \beta \lambda \rho}&=&s_4(\xi)(\hat{k}_F)_{\alpha \sigma \beta \tau}(\hat{k}_F)_{\lambda \, \, \, \rho}^{\, \, \, \, \sigma \, \, \, \tau}+s_5(\xi)(k_F)_{\alpha \beta} (k_F)_{\lambda \rho}\, ,
\end{eqnarray}
These finite contributions to the vacuum polarization arise through the following dimension-six interactions:
\begin{eqnarray}
	(\hat{k}_F)^{\alpha \omega \tau \sigma}(\hat{k}_F)^\beta_{\, \, \, \, \omega \tau \sigma}{\cal D}^{ab}_\alpha F^b_{\lambda \rho} {\cal D}^{ac}_\beta F^{c\lambda \rho}\,\, , (k_F)^{\alpha \sigma}(k_F)^\beta_{\, \, \, \, \sigma}{\cal D}^{ab}_\alpha F^b_{\lambda \rho} {\cal D}^{ac}_\beta F^{c\lambda \rho}\, , (\hat{k}_F)^{\alpha \tau \beta \sigma}(k_F)_{\tau \sigma}{\cal D}^{ab}_\alpha F^b_{\lambda \rho} {\cal D}^{ac}_\beta F^{c\lambda \rho}\, ,
\end{eqnarray}
and from the following dimension-eight interactions:
\begin{eqnarray}
	(\hat{k}_F)^{\alpha \omega \beta \eta}(\hat{k}_F)_{\sigma \omega \tau\eta}{\cal D}^{ab}_\alpha {\cal D}^{bc\sigma} F^c_{\lambda \rho} {\cal D}^{ad}_\beta  {\cal D}^{de\tau} F^{e\lambda \rho}\,\, , \, (k_F)^{\alpha \beta} (k_F)^{\sigma \tau}{\cal D}^{ab}_\alpha {\cal D}^{bc}_\sigma F^c_{\lambda \rho} {\cal D}^{ad}_\beta  {\cal D}^{de}_\tau F^{e\lambda \rho}\,
\end{eqnarray}
All the form factors $s_i(\xi)$ $(i=1,\cdots,5)$ are free of UV divergences but gauge-dependent. They are given in Appendix~\ref{AP}.\\

For clarity and later use, it is convenient to summarize the most outstanding results of the second order CPT-even  contribution. To this order, a Riemman-like UV divergent interaction $(k^{(2)}_F)_{\mu \lambda \nu \rho}$ arises. Once expressed into its irreducible pieces, UV divergent contributions are induced on Weyl-like, Ricci-like, and usual ($P_{\mu \nu}$) Lorentz structures. The Weyl-like contribution is given in Eq.~(\ref{Gamma2hat}) through the $(\hat{k}^{(2)}_F)_{\mu \lambda \nu \rho}$ tensor of Eq.~(\ref{Weyl2}). The Ricci-like contribution is incorporated into the $(\tilde{k}^{(2)}_F)_{\mu \nu }$ tensor given by Eq.~(\ref{Ricci2}), which also contains contributions from other sources. One of these contributions is the one characterized by the form factor $g_1(q^2,\xi)$, which does not involve the Ricci-like tree-level structure $(k_F)_{\mu \nu}$ but it emerges from a contraction between two Weyl-like structures. This means that, to second order, renormalization theory requires the presence of a non-zero Ricci-like tensor at the tree level.\\

To second order, the vacuum polarization function associated with the usual $P_{\mu \nu}$ Lorentz tensor also receives contributions from the scalar curvature induced by the irreducible parts of the $(k^{(2)}_F)_{\mu \lambda \nu \rho}$ Riemann-like tensor, but there are also contributions from other sources. All contributions are collected in the vacuum polarization function $\Pi^{E(2)}$ given in Eq.~(\ref{VPF2}). Below, we will study the implications of this contributions on the beta function $\beta(g)$. \\

There are no IR divergences arising from this sector. The amplitude (\ref{SO}) satisfies the Ward identity:
\begin{equation}
q^\mu \Pi^{E(2)}_{\mu \nu}(q,\xi)=q^\nu \Pi^{E(2)}_{\mu \nu}(q,\xi)=0\, ,
\end{equation}
which shows us that there is gauge invariance.

\subsubsection{CPT-Odd contribution}
The second order CPT-Odd contribution can be written as follows:
\begin{equation}\label{PiO2}
\Pi^{O(2)}_{\mu \nu}(q,\xi)=\Pi^{O(2)}(q^2,\xi)P_{\mu \nu}(q)+t_3(q^2,\xi)\frac{\Gamma^{AF}_{\mu \nu}(q)}{q^2}\, ,
\end{equation}
where
\begin{equation}
\label{PiO2}
\Pi^{O(2)}(q^2,\xi)=t_1(q^2,\xi)\frac{(k_{AF})^2}{q^2}+t_2(q^2,\xi)\frac{q^\alpha q^\beta (k_{AF})_\alpha(k_{AF})_\beta}{(q^2)^2} \, .
\end{equation}
 In the above expressions, $(k_{AF})^2=(k_{AF})_\kappa (k_{AF})^\kappa$ and $\Gamma^{AF}_{\mu \nu}(q)$ is given by:
\begin{equation}\label{RicciO2}
\Gamma^{AF}_{\mu \nu}(q)=\left(q^2\delta_\mu^\alpha\delta_\nu ^\beta+q^\alpha q^\beta g_{\mu \nu}- q_\mu q^\alpha \delta_\nu^ \beta-q_\nu q^\alpha  \delta_\mu^\beta\right)(k_{AF})_\alpha (k_{AF})_\beta \, .
\end{equation}
 Notice that the vacuum polarization receives a contribution from a dimension-six interaction, which has the form
\begin{equation}
	(k_{AF})^\alpha (k_{AF})^\beta {\cal D}^{ab}_{\alpha}F^b_{\lambda \rho}{\cal D}^{ac}_{\beta}F^{c\lambda\rho}\, .
\end{equation}
The $t_i(q^2,\xi)$ $(i=1,2,3)$ functions are free of UV divergences, but they have IR divergences. We have regulated these divergences by introducing a fictitious mass $m$ for the gauge fields. These form factors are presented in Appendix~\ref{AP}.\\

In Ref.\cite{MassT}, the emergence in this sector of a mass term proportional to $(k_{AF})^2 g_{\mu \nu}$ was suggested. In that work, the authors arrive at such conclusion within the framework of the Landau gauge. In our case, such mass term cannot arise because our amplitude is gauge-invariant, that is, it satisfies the Ward identity
\begin{equation}
q^\mu\Pi^{O(2)}_{\mu \nu}(q,\xi)=q^\nu\Pi^{O(2)}_{\mu \nu}(q,\xi)=0\, .
\end{equation}
This result is a consequence of the use of the BFM-gauge, which preserves gauge invariance.

\subsubsection{CPT-Even and CPT-Odd interference}
To second order, interference effects between CPT-Even and CPT-Odd terms are generated. Since the mSME is renormalizable, such effects must be free of UV divergences. Therefore, the calculation of these effects is important to test the internal consistency of the model. The corresponding amplitude is given by:
\begin{equation}\label{PiEO}
\Pi^{EO(2)}_{\mu \nu}(q,\xi)=-i \left( \frac{(k_{AF})^\kappa q^\lambda}{q^2}\Gamma^{EO}_{\kappa \lambda \mu \nu}+
\frac{(k_{AF})^\kappa q^\lambda q^\alpha q^\beta}{(q^2)^2}\Gamma^{EO}_{\kappa \lambda \alpha \beta\mu \nu}\right)\, ,
\end{equation}
where
\begin{eqnarray}
\Gamma^{EO}_{\kappa \lambda \mu \nu}&=&\left(q^2\delta^\alpha _\mu \delta^\beta_\nu-q_\mu q^\alpha \delta^\beta_\nu+q_\nu q^\alpha \delta^\beta_\mu \right)(k_{EO})_{\kappa \lambda \alpha \beta}\, , \label{GammaEO5}\\
\Gamma^{EO}_{\kappa \lambda \alpha \beta \mu \nu}&=&\left(q^2\delta^\sigma _\mu \delta^\tau_\nu-q_\mu q^\sigma \delta^\tau_\nu+q_\nu q^\sigma \delta^\tau_\mu \right)(k_{EO})_{\kappa \lambda \alpha \beta \sigma \tau}\, ,\label{GammaEO7}\
\end{eqnarray}
with the $(k_{EO})_{\kappa \lambda \alpha \beta}$ and $(k_{EO})_{\kappa \lambda \alpha \beta \sigma \tau}$ tensors given by:
\begin{eqnarray}
(k_{EO})_{\kappa \lambda \mu \nu}&=&\eta_1(\xi)\left(\delta^\alpha_\mu \epsilon_{\sigma \tau \kappa \nu}- \delta^\alpha_\nu \epsilon_{\sigma \tau \kappa \mu}\right)(\hat{k}_F)_{\alpha \lambda}^{\, \, \, \, \, \, \, \sigma \tau}+\eta_2(\xi)\left(\delta^{\alpha}_\mu \epsilon_{\sigma \kappa \lambda \nu} -
\delta^{\alpha}_\nu \epsilon_{\sigma \kappa \lambda \mu} \right)(k_F)_\alpha^{\, \, \, \sigma} \nonumber \\
&&+\eta_3(\xi)(k_F)_{\lambda}^{\, \, \,  \sigma}\epsilon_{\sigma \kappa \mu \nu}\, ,
\end{eqnarray}
\begin{equation}
	(k_{EO})_{\kappa \lambda \alpha \beta \mu \nu}=\eta_4(\xi)\left(\delta^\tau_\nu \epsilon_{\sigma \kappa \lambda \mu}-\delta^\tau_\mu
	\epsilon_{\sigma \kappa \lambda \nu} \right)(\hat{k}_F)_{\tau \alpha \beta}^{\, \, \, \, \, \, \, \, \, \, \sigma}+\eta_5(\xi)\epsilon_{\kappa \lambda \mu \nu}(k_F)_{\alpha \beta}\, .
\end{equation}
Notice that these tensors have the following symmetries: $(k_{EO})_{\kappa \lambda \mu \nu}=-(k_{EO})_{\kappa \lambda \nu \mu}$ and $(k_{EO})_{\kappa \lambda \alpha \beta \mu \nu}=-(k_{EO})_{\kappa  \lambda \alpha \beta \nu \mu}$. The antisymmetry of these tensors in the pair of indices $\mu$ and $\nu$ does not mean that the amplitude $\Pi^{EO(2)}_{\mu \nu}$ is antisymmetric under the interchange of indices $\mu \leftrightarrow \nu$, since this interchange must be realized together with the change $q \rightarrow-q$, which leads to a symmetric amplitude. On the other hand, the form factors $\eta_i(\xi)$ are free of both UV and IR divergencies, but they are gauge-dependent. They are given in Appendix~\ref{AP}.\\

The amplitude~(\ref{PiEO}) corresponds to dimension-five and dimension-seven interactions. The dimension-five interactions associated with the $(k_{EO})_{\kappa \lambda \mu \nu}$ tensor have the form
\begin{eqnarray}
	&&(k_{AF})^{\kappa}(\hat{k_{F}})^{\alpha\lambda\sigma\tau}\epsilon_{\sigma\tau\kappa\beta}\left(F^{a}_{\alpha\gamma}{\cal D}^{ab}_{\lambda}F^{b\gamma\beta} -F^{a}_{\beta\gamma}{\cal D}^{ab}_{\lambda}F^{b\gamma\alpha}\right) \, , \,\,  \nonumber\\	 &&(k_{AF})^{\kappa}(k_{F})^{\alpha\sigma}\epsilon_{\sigma\kappa\lambda\beta}\left(F^{a}_{\alpha\gamma}{\cal D}^{ab\lambda}F^{b\gamma\beta} -F^{a}_{\beta\gamma}{\cal D}^{ab\lambda}F^{b\gamma\alpha}\right)\, , \,\, \nonumber\\
	&&(k_{AF})^{\kappa}(k_{F})^{\lambda\sigma}\epsilon_{\sigma\kappa\alpha\beta}F^{a\alpha}{}_{\gamma}{\cal D}^{ab}_{\lambda}F^{b\gamma\beta} \,\, ,
\end{eqnarray}
while the dimension-seven interactions associated with the $(k_{EO})_{\kappa \lambda \alpha \beta \mu \nu}$ tensor arise from
\begin{equation}
(k_{AF})_{\kappa} \epsilon^{\sigma\kappa\lambda\mu}(\hat{k}_{F})^{\nu\alpha\beta}{}_{\sigma}  \left(  {\cal D}^{ab}_{\alpha}F^{b}_{\mu\gamma}{\cal D}^{ad}_{\beta}{\cal D}^{de}_{\lambda}F^{e\gamma}{}_{\nu}-{\cal D}^{ab}_{\alpha}F^{b}_{\nu\gamma}{\cal D}^{ad}_{\beta}{\cal D}^{de}_{\lambda}F^{e\gamma}{}_{\mu}\right)\,\, ,\,\, (k_{AF})_{\kappa}\epsilon^{\kappa\lambda\mu\nu}(k_{F})^{\alpha\beta}{\cal D}^{ab}_{\alpha}F^{b}_{\mu\gamma}{\cal D}^{ad}_{\beta}{\cal D}^{de}_{\lambda}F^{e\gamma}{}_{\nu}\: .
\end{equation}
The amplitude is gauge invariant since it satisfies the Ward identity
\begin{equation}
	q^\mu \Pi^{EO(2)}_{\mu \nu}(q,\xi)=q^\nu \Pi^{EO(2)}_{\mu \nu}(q,\xi)=0\, .
\end{equation}

\section{Renormalization and beta functions}
\label{RBF}
One of the main advantages of using the BFM-gauge is that it preserves gauge-invariance with respect to the background gauge fields. This means that the usual beta function, $\beta(g)$, can be derived from the renormalized two point $\Pi^{YME ab}_{\mu \nu}(q)$ vertex function. The purposes of this section is to derive such beta function, as well as those associated with the LV coefficients. To analyze the decoupling or non-decoupling nature of the new physics effects, we will carry out our study of the usual beta function by implementing two renormalization schemes, namely a mass-independent scheme and a mass-dependent scheme.\\

Up to second order in the LV coefficients, the vacuum polarization function is given by Eqs.~(\ref{Pi}), (\ref{PiE1}), (\ref{PiE2}), (\ref{VPF2}), and (\ref{PiO2}):
\begin{equation}
\label{VPF}
\Pi_{YME}(q^2,\xi)=\Pi(q^2,\xi)+\Pi^{E(1)}(q^2,\xi)+\Pi^{E(2)}(q^2,\xi)+\Pi^{O(2)}(q^2,\xi)-\delta_A\, ,
\end{equation}
where the counterterm has be included. Remember that only $\Pi(q^2,\xi)$ and the form factors $h_1(q^2,\xi)$ and $h_2(q^2,\xi)$ of $\Pi^{E(2)}(q^2,\xi)$ have UV divergences.

\subsection{Mass-independent scheme}
We will use the $\overline{MS}$ scheme, in which the counterterm is defined by a term of the form $c\Delta$, being $c$ a constant independent of the external moment and $\Delta$ the divergent quantity given in Eq.~(\ref{Del}).\\

\subsubsection{The $\beta_{YME}(g)$ function}
In a mass-independent renormalization scheme, the beta function is given by:
\begin{equation}
\label{betadef}
\beta^{\overline{MS}}_{YME}(g)=-\frac{1}{2}g^2\frac{\partial Z^{(1)}_A}{\partial g}\, ,
\end{equation}
where $Z^{(1)}_A$ is the coefficient of the simple pole of $Z_A$.\\

The renormalization factor $Z_A$ arises from Eq.~(\ref{VPF}), so in the $\overline{MS}$ scheme, it is given by:
\begin{equation}
\label{Za}
Z_A=1+\frac{g^2C_2(G)}{(4\pi)^2}\frac{1}{\epsilon}\left[\frac{11}{3}+\frac{3}{8}(\hat{k}_F)^2+\frac{25}{18}(k_F)^2\right]\, .
\end{equation}

 Then, from Eqs.~(\ref{betadef}) and (\ref{Za}), the beta function becomes:
\begin{equation}
\label{betaMS}
\beta^{\overline{MS}}_{YME}(g)=\beta(g)\left[1+\frac{9}{88}(\hat{k}_F)^2+\frac{25}{66}(k_F)^2\right]\, ,
\end{equation}
where $\beta(g)$ is the usual beta function of a pure (without matter fields) Yang-Mills theory. Of course, our result reduces to the usual one when the $(\hat{k}_F)_{\mu \lambda \nu \rho}$ and $(k_F)_{\mu \nu}$ tensors are identically zero, but it can also happen in presence of new physics if $\frac{9}{88}(\hat{k}_F)^2+\frac{25}{66}(k_F)^2=0$, which is possible due to the spacetime metric. Other possibilities are: $\frac{9}{88}(\hat{k}_F)^2+\frac{25}{66}(k_F)^2>0$ or $\frac{9}{88}(\hat{k}_F)^2+\frac{25}{66}(k_F)^2<0$. In the first case the phenomenon of asymptotic freedom is reinforced, while in the second case this phenomenon is weakened.\\

In the $\overline{MS}$ scheme, the tensor counterterms of the CPT-Even sector are determined by the following conditions:
\begin{eqnarray}
\label{RC1}
Z_A(Z_{\hat{k}_F})^{\mu \nu \alpha \beta}_{\, \, \, \, \, \,\, \, \,\, \, \, \, \lambda \rho \gamma \sigma}(\hat{k}_F)^{\lambda \rho \gamma \sigma}-
(\hat{k}_F)^{\mu \nu \alpha \beta}+f^{\Delta}_1(\hat{k}_F)^{\mu \nu \alpha \beta}+(\hat{k}^{(2)}_F)^{\mu \nu \alpha \beta}_\Delta &=&0 \, ,\\
\label{RC2}
Z_A(Z_{k_F})^{\mu \alpha}_{\, \, \, \, \, \,\,\lambda \rho}(k_F)^{\lambda \rho}-(k_F)^{\mu \alpha}+f^\Delta_2(k_F)^{\mu \alpha}+(\tilde{k}^{(2)}_F)^{\mu \alpha}_\Delta &=&0\, ,
\end{eqnarray}
where $f^{\Delta}_1$ and $f^{\Delta}_2$ are the UV divergent parts of the $f_1(q^2,\xi)$ and $f_2(q^2,\xi)$ form factors, whereas $(\hat{k}^{(2)}_F)^{\mu \nu \alpha \beta}_\Delta$ and $(\tilde{k}^{(2)}_F)^{\mu \alpha}_\Delta$ are the UV divergent parts of the $(\hat{k}^{(2)}_F)^{\mu \nu \alpha \beta}$ and $(\tilde{k}^{(2)}_F)^{\mu \alpha}$ tensors.\\

On the other hand, we have seen that the CPT-Odd sector does not generate UV divergences, so in the  $\overline{MS}$ scheme the counterterm  associated with this sector is null:
\begin{equation}
\label{RC3}
Z_A(Z_{k_{AF}})^\kappa_{\, \, \, \lambda}(k_{AF})^\lambda-(k_{AF})^\kappa=0\, .
\end{equation}

Now, to first order in $\alpha_g=\frac{g^2}{4\pi}$ and up to second order in the LV coefficient, from (\ref{Za}) we can write:
\begin{equation}
\label{ZaI}
Z_A^{-1}=1-\frac{\alpha_g C_2(G)}{4\pi}\frac{1}{\epsilon}\left[\frac{11}{3}+\frac{3}{8}(\hat{k}_F)^2+\frac{25}{18}(k_F)^2\right]\, .
\end{equation}
So, working always up to this order, Eqs.~(\ref{RC1}), (\ref{RC2}), and (\ref{RC3}) become:
\begin{eqnarray}
\label{FR1}
(Z_{\hat{k}_F})^{\mu \nu \alpha \beta}_{\, \, \, \, \, \,\, \, \,\, \, \, \, \lambda \rho \gamma \sigma}(\hat{k}_F)^{\lambda \rho \gamma \sigma}&=&(\hat{k}_F)^{\mu \nu \alpha \beta}+\frac{\alpha_g C_2(G)}{4\pi}\frac{1}{\epsilon}\left[\frac{7}{3}(\hat{k}_F)^{\mu \nu \alpha \beta}+\frac{3}{2}(\hat{k}^{(2)}_W)^{\mu \nu \alpha \beta} +\frac{7}{6}(\hat{k}^{(2)}_R)^{\mu \nu \alpha \beta}  \right]\, , \\
\label{FR2}
(Z_{k_F})^{\mu \alpha}_{\, \, \, \, \, \,\,\lambda \rho}(k_F)^{\lambda \rho}&=&(k_F)^{\mu \alpha}+\frac{\alpha_g C_2(G)}{4\pi}\frac{1}{\epsilon} \Bigg\{ \frac{1}{2}(\hat{k}_F)^{\mu \lambda \rho \sigma}(\hat{k}_F)^\alpha_{\, \, \, \lambda \rho \sigma}\nonumber \\
&&+\frac{7}{3}\left[(\hat{k}_F)^{\mu \lambda \alpha \sigma}(k_F)_{\lambda \alpha}+ (k_F)^{\mu \lambda}(k_F)^\alpha_{\, \, \, \lambda}\right]\Bigg \}\, ,\\
\label{FR3}
(Z_{k_{AF}})^\kappa_{\, \, \, \lambda}(k_{AF})^\lambda &=&(k_{AF})^\kappa-\frac{11}{3}\frac{\alpha_g C_2(G)}{4\pi}\frac{1}{\epsilon}(k_{AF})^\kappa \, .
\end{eqnarray}
Some comments are in order here. From Eq.~(\ref{FR2}), it can be seen that the UV pole of the Ricci-like renormalization factor does not depend linearly on the LV coefficients, which is due to an exact cancellation with the expression for $Z^{-1}_A$. Also, notice that, although the CPT-Odd contribution is free of UV divergences, the corresponding renormalization factor is UV divergent through the $Z^{-1}_A$ renormalization factor.\\

In the $\overline{MS}$ scheme, the beta functions can be determined from the simple pole of the corresponding renormalization factor, that is,
\begin{eqnarray}
(\beta_{\hat{k}_F})_{\mu \nu \alpha \beta}&=&2\alpha_g\frac{d(a^{\hat{k}_F}_1)_{\mu \nu \alpha \beta}}{d\alpha_g}\, , \\
(\beta_{k_F})_{\mu \alpha}&=&2\alpha_g\frac{d(a^{k_F}_1)_{\mu \alpha }}{d\alpha_g}\, ,\\
(\beta_{k_{AF}})_{\kappa}&=&2\alpha_g\frac{d(a^{k_{AF}}_1)_{\kappa}}{d\alpha_g}\, ,
\end{eqnarray}
where $(a^{\hat{k}_F}_1)_{\mu \nu \alpha \beta}$, $(a^{k_F}_1)_{\mu \alpha }$, and $(a^{k_{AF}}_1)_{\kappa}$ are the coefficients of the simple pole in Eqs.~(\ref{FR1}), (\ref{FR2}), and (\ref{FR3}). Then, the beta functions are given by:
\begin{eqnarray}
(\beta_{\hat{k}_F})^{\mu \nu \alpha \beta}&=&\frac{\alpha_g C_2(G)}{4\pi}\left[\frac{14}{3}(\hat{k}_F)^{\mu \nu \alpha \beta}+3(\hat{k}^{(2)}_W)^{\mu \nu \alpha \beta}+\frac{7}{3} (\hat{k}^{(2)}_R)^{\mu \nu \alpha \beta}\right]\, , \\
(\beta_{k_F})^{\mu \alpha}&=&\frac{\alpha_g C_2(G)}{4\pi}\left\{(\hat{k}_F)^{\mu \lambda \rho \sigma}(\hat{k}_F)^\alpha_{\, \, \, \lambda \rho\sigma} +\frac{14}{3}\left[(k_F)^{\mu \lambda}(k_F)^\alpha_{\, \, \, \lambda}+(\hat{k}_F)^{\mu \lambda \alpha \rho}(k_F)_{\lambda \rho}\right]\right\}\, , \\
(\beta_{k_{AF}})^\kappa&=&-\frac{22 \alpha_g C_2(G)}{3(4\pi)}(k_{AF})^\kappa\, .
\end{eqnarray}
Notice that the beta function associated with the Ricci-like $(k_F)_{\mu \alpha}$ tensor is generated up to second-order in the LV coefficients. We also note that if only first-order contributions are maintained, our results reduce to those given in Ref.~\cite{YMER}.

\subsection{Mass-dependent scheme}
We now proceed to study the structure of the beta function $\beta_{YME}(g)$ in a mass-dependent renormalization scheme. The idea is to analyze the decoupling or non-decoupling nature of the new physics effects. In this scheme, the renormalization condition is:
\begin{equation}
\Pi_{YME}(q^2=-\mu^2,\xi)=0\, ,
\end{equation}
so, the counterterm is given by:
\begin{equation}
\delta_A=\left[\Pi(q^2,\xi)+\Pi^{E(1)}(q^2,\xi)+\Pi^{E(2)}(q^2,\xi)+\Pi^{O(2)}(q^2,\xi)  \right]_{q^2=-\mu^2}\, .
\end{equation}
The beta function is given by:
\begin{eqnarray}
\beta^{(\mu^2)}_{YME}(g)&=&g\mu^2 \frac{\partial Z_A(-\mu^2)}{\partial \mu^2}\nonumber \\
&=&gq^2\frac{\partial Z_A(q^2)}{\partial q^2}\Big|_{q^2=-\mu^2}\nonumber \\
&=&\frac{g}{2}q^\alpha \frac{\partial Z_A(q^2)}{\partial q^{\alpha}}\Big|_{q^2=-\mu^2}\, .
\end{eqnarray}
Now, using the following identity:
\begin{equation}
q^\alpha \frac{\partial}{\partial q^\alpha}\left(\frac{q^{\alpha_1} \cdots q^{\alpha_{2n}}}{(q^2)^n}\right)=0\, ,
\end{equation}
it can be shown that $\Pi^{E(1)}(q^2,\xi)$ and the finite part of $\Pi^{E(2)}(q^2,\xi)$ do not contribute to the beta function. Therefore,
\begin{equation}
\beta^{(\mu^2)}_{YME}(g)=g\left[\mu^2 \frac{\partial \Pi(-\mu^2,\xi)}{\partial \mu^2}+ \mu^2 \frac{\partial \Pi^{E(2)}_{Div}(-\mu^2,\xi)}{\partial \mu^2} +
\frac{1}{2}q^\alpha \frac{\partial \Pi^{O(2)}(q,\xi)}{\partial q^\alpha}\Big|_{q^2=-\mu^2}\right]\, ,
\end{equation}
that is
\begin{eqnarray}
\label{betamu}
\beta^{(\mu^2)}_{YME}(g)&=&\beta(g)\Bigg\{1+\frac{9}{88}(\hat{k}_F)^2+\frac{25}{66}(k_F)^2\nonumber \\
&&-\frac{3}{22}\frac{(k_{AF})^2}{\mu^2}\left[(\xi+4)\log\left(\frac{m^2}{\mu^2}\right)+\frac{\xi}{2}(4-\xi)+\frac{3}{2}  \right]\nonumber \\
&&-\frac{3}{22}\frac{(\bar{k}_{AF})^2}{(\mu^2)^2}\left[(\xi+3)\log\left(\frac{m^2}{\mu^2}\right) +\frac{1}{2}(1+\xi)(5-\xi)\right] \Bigg\}\, ,
\end{eqnarray}
where we have introduced the following definition $\bar{k}_{AF}=q^\alpha (k_{AF})_\alpha$. Notice that this quantity is invariant under Lorentz observer transformations, but not under particle transformations.\\

The expression for the beta function given in (\ref{betamu}) has some interesting ingredients related to the CPT-Odd contribution that need to be clarified. The contribution of the CPT-Odd sector to the beta function in this mass-dependent renormalization scheme is not surprising, since the Lorentz coefficient $k_{AF}$ has units of mass. This behavior of the beta function in this type of renormalization schemes is already observed in the case of a Yang-Mills theory with a fermion sector given in an $\mathbf{r}$ representation $\psi$ of the $SU(N)$ group. In this case, the usual beta function is given by:
\begin{equation}
\beta^{(\mu^2)}(g)=\frac{g^3}{(4\pi)^2}\left[-\frac{11}{3} C_2(G)+8C(\mathbf{r})\int^1_0dx \frac{\mu^2 x^2(1-x)^2}{m^2_\psi +\mu^2 x(1-x)} \right]\, ,
\end{equation}
which reduces to the well-known result of the $\overline{MS}$ scheme in the $\mu^2\gg m^2_\psi$ limit:
\begin{equation}
\beta^{\overline{MS}}(g)=\frac{g^3}{(4\pi)^2}\left[-\frac{11}{3} C_2(G) +\frac{4}{3}C(\mathbf{r})\right]\, .
\end{equation}

In our case, we recover the $\overline{MS}$-scheme result (\ref{betaMS}) in the $\mu^2 \gg (k_{AF})^2$ and $(\mu^2)^2 \gg (\bar{k}_{AF})^2$ limits. \\

On the other hand, the gauge dependence of $\beta^{(\mu^2)}_{YME}(g)$ is puzzling because it indicates that it cannot be a physical quantity, at least at energies of the order of $k_{AF}$. This gauge-dependence of the beta function arises as a consequence of the fact that $(k_{AF})_\mu$ is a gauge coefficient in the sense that it is structurally linked to the gauge sector of the theory. Indeed, this behavior is not exclusive of gauge sectors that are odd under CPT transformations. Gauge-dependence of the beta function in a $\mu$-scheme can also arise in gauge theories with spontaneous symmetry breaking. For instance, in the SM, the contribution of the $W^\pm_\mu$ weak gauge boson to the electromagnetic beta function is also gauge-dependent in a $\mu$-scheme.\\

From the above considerations, we can conclude that our result (\ref{betamu}) for the beta function is within what can be expected in a conventional quantum field theory. The ingredient in our result that is not common to conventional field theories is the presence of IR divergences. As it was commented in the introduction, IR divergences in the context of the mSME also arise in physical quantities as anomalous magnetic moments~\cite{OP2}. The presence of this type of divergences in physical observables seems to be an undesirable characteristic of the mSME, which will require the implementation of complicated cancelation mechanisms~\cite{OP2}.

\section{Summary}
\label{C}
In this work, we have studied the one-loop structure of the Yang-Mills Extension without matter fields. This renormalizable version of the theory has both a CPT-Even sector and a CPT-Odd sector. The CPT-Even sector is characterized by a dimensionless Riemann-like tensor, which,  for renormalization purposes, is decomposed into its irreducibles pieces, namely a Weyl-like $(\hat{k}_F)_{\mu \nu \lambda \rho}$ tensor, a Ricci-like $(k_F)_{\mu \nu}$ tensor, and a $\bar{k}_F$ scalar, analogous of the scalar curvature. The $\bar{k}_F$ coefficient can be removed from the theory through a rescaling  of the gauge field $A^a_\mu$, the coupling constant $g$, and the $(\hat{k}_F)_{\mu \nu \lambda \rho}$ and $(k_F)_{\mu \nu}$ tensors. As far as the CPT-Odd sector is concerned, it is characterized by a $(k_{AF})_\mu$ vector, which has units of mass. These coefficients transform under observer Lorentz transformations but not under particle Lorentz transformations.\\

In order to study Lorentz violation effects on the beta function associated to the coupling constant $g$ and also to study the internal consistency of the theory, effects up to second order in the Lorentz coefficients were considered. To simplify the analysis, the BFM-gauge was introduced, which allows us to derive the diverse beta functions from the $A^a_\mu A^b_\nu$ two-point vertex function. The use of the BFM-gauge greatly simplified the one-loop calculations, which, by the way, in a linear $R_\xi$-gauge could constitute a formidable challenge. In this gauge, the renormalization constants of the gauge field $A^a_\mu$ and the coupling constant $g$ are related. We have focused in a special way on the usual beta function, which receives its first contribution up to the second order in the Lorentz violating coefficients. We have carried out our calculations maintaining the gauge parameter $\xi$, which has allowed us to study the gauge-dependence or gauge-independence of the various magnitudes of interest. In the literature, the Ricci-like $(k_F)_{\mu \nu}$ tensor has been considered equal to zero, but we have found that its presence is required by renormalization theory when second-order effects are considered, since, to this order, certain contraction of indices in the squared of the Weyl-like tensor leads to a Ricci-like tensor. Exact results are presented, including second-order interference effects between the CPT-Even and CPT-Odd sectors. These interference effects are free of UV divergences, which is in agreement with renormalization theory. Effects free of UV divergences characterized by dimension-six and dimension-eight interactions are induced. Our result for the two-point $A^a_\mu A^b_\nu$ vertex function satisfies the Ward Identity, which means that it respects gauge-invariance, so a mass term proportional to the CPT-Odd dimensionfull parameter $(k_{AF})^2$ is not generated, as has been suggested in the literature.\\

To first order, the CPT-Even sector induces UV divergent amplitudes proportional to the Weyl-like $(\hat{k}_F)_{\mu \nu \lambda \rho}$ tensor and to Ricci-like $(k_F)_{\mu \nu}$ tensor, as well as a finite contribution proportional to the usual vacuum polarization tensor structure $P_{\mu \nu}$.
On the other hand, the CPT-Odd sector induces a contribution free of UV divergences, which is proportional to the Lorentz structure introduced at the tree level.\\

Second order effects are much more complicated. In the CPT-Even sector, UV divergent contributions proportional to the Weyl-like tensor $(\hat{k}^{(2)}_F)_{\mu \lambda \nu \rho}$, the Ricci-like tensor $(k^{(2)}_F)_{\mu \nu}$, and to the vacuum polarization tensor are generated. Contributions free of UV divergences are also generated by this sector. Contributions free of UV divergences of Ricci type are induced by dimension-six and dimension-eight interactions characterized by the $(\tilde{k}^{(2)}_F)_{\mu \nu \lambda \rho}$ and $(\tilde{k}^{(2)}_F)_{\mu \nu \lambda \rho \sigma \tau}$ tensors, Eqs.~(\ref{Ricci6}) and (\ref{Ricci8}), respectively. Also, finite contributions to the vacuum polarization are induced by dimension-six and dimension-eight interactions. These contributions are characterized by the $(k_1)_{\alpha \beta}$ and $(k_2)_{\alpha \beta \lambda \rho}$ tensors given in Eqs.~(\ref{k1}) and (\ref{k2}), respectively. These interactions of dimension higher than four involve products between the Weyl-like $(\hat{k}_F)_{\mu \nu \lambda \rho}$ and the Ricci-like $(k_F)_{\mu \nu}$ tensors in diverse combinations.  On the other hand, the CPT-Odd sector induces contributions free of UV divergences, but they are IR divergent. These contributions are proportional to the vacuum polarization $P_{\mu \nu}$ tensor and a interaction proportional to the  symmetric  $(k_{AF})_\mu (k_{AF})_\nu$ tensor,  Eq.~(\ref{PiO2})\\

The interference between the CPT-Even and CPT-Odd sectors is generated through dimension-five and dimension-seven interactions, which involve products of the $(k_{AF})_\mu$ vector with the Weyl-like $(\hat{k}_{F})_{\mu\alpha\nu\beta}$ or the Ricci-like $(k_F)_{\alpha \beta}$ tensors. Although this result is free of both UV and IR divergences, it is gauge-dependent.\\

Second order corrections to the beta functions arising from the Weyl-like $(\hat{k}_{F})_{\mu\nu\alpha\beta}$  and Ricci-like $(k_{F})_{\mu\alpha}$ tensors were included. Nevertheless, the beta function $(\beta_{k_{F}})_{\mu\alpha}$ is generated up to second order in the LV parameters. On another hand, the beta function associated with the coupling constant $g$ was studied in both a mass-independent and in a mass-dependent renormalization schemes. We found that, in the $\overline{MS}$ scheme, the beta function only receives contributions from the CPT-Even sector, which are proportional to the $(\hat{k}_F)^2$ and $(k_F)^2$ scalars. On the other hand, in the $\mu^2$-scheme, the beta function receives, in addition, contributions from the CPT-Odd sector, which arise as a consequence of the fact that the $(k_{AF})_\alpha$ coefficient has units of mass. This result reduces to the one of the $\overline{MS}$ scheme in the $(k_{AF})^2\ll \mu^2$ and $(\bar{k}_{AF})^2\ll (\mu^2)^2$ limits. The contribution of this dimensional parameter is not surprising, since the same behavior is observed in conventional theories when the masses of the particles are not neglected. Although such contribution is gauge-dependent, it agrees with the fact that the  $(k_{AF})_\alpha$ coefficient is a gauge parameter, in the sense that it is linked to the gauge sector. Actually, the same behavior is observed in theories with spontaneous symmetry breaking. In this class of theories, contributions to the beta function arising from massive gauge bosons are gauge-dependent in a $\mu$-scheme. All of this is within what is expected in the context of a conventional quantum field theory. What is intriguing is the presence of IR divergences, a phenomenon that, as has been shown in other contexts, seems to be exclusive of theories with Lorentz violation. In general, the presence of IR divergences in well-defined observables in conventional theories, such as, for example, anomalous magnetic moments, will require considering, in each case, a more complete process, without a conventional analogue, that allows its cancelation.

\appendix

\section{Form factors}
\label{AP}
In this appendix, we list the form factors generated from loop calculations. These loop amplitudes are exact and have been obtained in the general BFM-gauge, so they depend on the gauge-parameter $\xi$.

\subsection{First order form factors}
The form factors that arise from first order calculations are given by:
\begin{eqnarray}
f_1(q^2,\xi)&=&-\frac{g^2C_2(G)}{(4\pi)^2}\left[6\left(\Delta-\log\left(-\frac{q^2}{\hat{\mu}^2}\right)\right)-\frac{1}{2}(1-\xi)(7+\xi)+10\right]\, ,\\
f_2(q^2,\xi)&=&-\frac{g^2C_2(G)}{(4\pi)^2}\left[\frac{11}{3}\left(\Delta-\log\left(-\frac{q^2}{\hat{\mu}^2}\right)\right)
-\frac{1}{2}(1-\xi)(5+\xi)+\frac{67}{9}\right]\, ,\\
f_3(\xi)&=&-\frac{g^2C_2(G)}{(4\pi)^2}\left(\xi-\frac{2}{3}\right)\, .
\end{eqnarray}

\subsection{Second order form factors}
These form factors are organized according to whether they emerge from the CPT-Even sector or from the CPT-Odd sector.

\subsubsection{Form factors of the CPT-Even sector}
Form factors characterizing the Riemann-like tensor $(k^{(2)}_F)_{\mu \lambda \nu \rho}$.
\begin{eqnarray}
l_1(q^2,\xi)&=&-\frac{g^2C_2(G)}{(4\pi)^2}\left[\frac{3}{2}\left(\Delta -\log\left(-\frac{q^2}{\hat{\mu}^2}\right)\right)+\frac{\xi}{2}+\frac{10}{3} \right]\, ,\\
l_2(q^2,\xi)&=&-\frac{g^2C_2(G)}{(4\pi)^2}\left[\frac{7}{6}\left(\Delta -\log\left(-\frac{q^2}{\hat{\mu}^2}\right)\right)-\frac{1}{8}\left(\xi^2-\frac{29}{9}\right) \right]\, .
\end{eqnarray}

The total Ricci-like contribution is characterized by the following form factors:
\begin{eqnarray}
g_1(q^2,\xi)&=&-\frac{g^2C_2(G)}{(4\pi)^2}\left[\frac{1}{2}\left(\Delta -\log\left(-\frac{q^2}{\hat{\mu}^2}\right)\right)+\frac{\xi}{8}+\frac{37}{24}\right]\, ,\\
g_2(q^2,\xi)&=&-\frac{g^2C_2(G)}{(4\pi)^2}\left[\frac{7}{3}\left(\Delta -\log\left(-\frac{q^2}{\hat{\mu}^2}\right)\right)+\frac{1}{8}\xi (1-3\xi)+\frac{89}{36}\right]\, , \\
g_3(q^2,\xi)&=&-\frac{g^2C_2(G)}{(4\pi)^2}\left[\frac{7}{3}\left(\Delta -\log\left(-\frac{q^2}{\hat{\mu}^2}\right)\right)+\frac{3}{4}\xi+\frac{149}{36}\right]\, .
\end{eqnarray}

The form factors associated with the usual contribution $P_{\mu \nu}$ are given by:
\begin{eqnarray}
h_1(q^2,\xi)&=&\frac{g^2C_2(G)}{(4\pi)^2}\left[\frac{3}{8}\left(\Delta-\log\left(-\frac{q^2}{\hat{\mu}^2}\right)\right)+\frac{\xi}{6}+\frac{7}{9}\right]\, , \\
h_2(q^2,\xi)&=&\frac{g^2C_2(G)}{(4\pi)^2}\left[\frac{25}{18}\left(\Delta-\log\left(-\frac{q^2}{\hat{\mu}^2}\right)\right)+\frac{9}{432}\xi(9-2\xi)
+\frac{985}{432}\right]\, .
\end{eqnarray}

The form factors characterizing the dimension-six and dimension-eight interactions are given by:
\begin{eqnarray}
	g_4(\xi)&=&\frac{g^2C_2(G)}{(4\pi)^2}\left(\frac{\xi}{4}+\frac{5}{12}\right)\, , \\
	g_5(\xi)&=&\frac{g^2C_2(G)}{(4\pi)^2}\left[\frac{\xi}{4}\left(\xi+3\right)-\frac{5}{3}\right]\, ,\\
	g_6(\xi)&=&\frac{g^2C_2(G)}{(4\pi)^2}\left[\frac{\xi}{4}\left(2\xi+3\right)-\frac{1}{4}\right]\, ,\\
	g_7(\xi)&=&-\frac{g^2C_2(G)}{(4\pi)^2}\left(\frac{5}{4}\right)\left(\xi +3\right)\, ,\\
	g_8(\xi)&=&\frac{g^2C_2(G)}{(4\pi)^2}\left[\frac{\xi}{2}\left(\xi+2\right)-\frac{41}{6}\right]\, ,\\
	g_9(\xi)&=&-\frac{g^2C_2(G)}{(4\pi)^2}\left[\frac{\xi}{2}\left(\xi+2\right)-\frac{11}{2}\right]\, .
\end{eqnarray}

\begin{eqnarray}
s_1(\xi)&=&\frac{g^2C_2(G)}{(4\pi)^2}\frac{3}{8}(1-\xi)\, ,\\
s_2(\xi)&=&\frac{g^2C_2(G)}{(4\pi)^2}\left[\frac{2}{3}+\frac{\xi}{4}(3-\xi)\right]\, , \\
s_3(\xi)&=&\frac{g^2C_2(G)}{(4\pi)^2}\left(\frac{3}{4}\xi -\frac{17}{12}\right)\, , \\
s_4(\xi)&=&\frac{g^2C_2(G)}{(4\pi)^2}\frac{1}{6}(1-3\xi)\, , \\
s_5(\xi)&=&\frac{g^2C_2(G)}{(4\pi)^2}\frac{1}{12}(1-3\xi)\,.
\end{eqnarray}

\subsubsection{Form factors of the CPT-Odd sector}
The form factors associated with the CPT-Odd sector are free of UV divergences, but they present IR divergences. These functions, which have been  regulated with a fictitious mass $m$ of the gauge fields, are given by:
\begin{eqnarray}
t_1(q^2,\xi)&=&\frac{g^2C_2(G)}{(4\pi)^2}\left[\frac{1}{2}(\xi +4)\log\left(-\frac{m^2}{q^2}\right)+\frac{1}{4}\xi(2-\xi)-\frac{5}{4} \right]\, ,\\
t_2(q^2,\xi)&=&\frac{g^2C_2(G)}{(4\pi)^2}\left[-\frac{1}{2}(\xi +3)\log\left(-\frac{m^2}{q^2}\right)+\frac{1}{4}(1-\xi)^2\right]\, ,\\
t_3(q^2,\xi)&=&\frac{g^2C_2(G)}{(4\pi)^2}(\xi-1)\left[\log\left(-\frac{m^2}{q^2}\right)+1\right]\, .
\end{eqnarray}

\subsubsection{Form factors from interference effects}
Interference effects between CPT-Even and CPT-Odd terms induce the following forms factors.
\begin{eqnarray}
\eta_1(\xi)&=&-\frac{g^2C_(G)}{(4\pi)^2}\frac{3}{4}(1+\xi)\, , \\
\eta_2(\xi)&=&-\frac{g^2C_(G)}{(4\pi)^2}\frac{1}{4}\left[\xi(\xi+1)+4\right]\, , \\
\eta_3(\xi)&=&\frac{g^2C_(G)}{(4\pi)^2}\frac{3}{2}(\xi+1)\, ,\\
\eta_4(\xi)&=&\frac{g^2C_(G)}{(4\pi)^2}\frac{1}{2}(\xi^2-5)\, ,\\
\eta_5(\xi)&=&\frac{g^2C_(G)}{(4\pi)^2}\frac{1}{2}(\xi^2-1)\, ,
\end{eqnarray}

\acknowledgments{We acknowledge financial support from CONACYT and
SNI (M\' exico).}

\end{document}